\documentclass[%
aip,
 jcp,
 amsmath,amssymb,
 reprint,%
]{revtex4-1}

\usepackage{graphicx}
\usepackage{dcolumn}
\usepackage{bm}

\usepackage[utf8]{inputenc}
\usepackage[T1]{fontenc}
\usepackage{mathptmx}
\usepackage{etoolbox}
\usepackage{bbm}
\usepackage{lipsum}
\usepackage{xcolor}
\usepackage{tikz}
\usepackage[section]{placeins}

\newcommand{\pol}{\text{pol}}
\newcommand{\stall}{\text{stall}}
\newcommand{\irr}{\text{irrev}}
\newcommand{\act}{\text{act}}
\newcommand{\KP}{\text{KP}}
\newcommand{\ina}{\text{in}}
\newcommand{\abs}{\text{com}}
\newcommand{\Gpol}{\Delta  G_{\pol}}

\newcommand{\citeTSA}{\cite{sahoo2021accuracy,song2016proofreading,song2020thermodynamic,li2019template,pigolotti2016protocols,sartori2015thermodynamics,wong2018energy,pineros2020kinetic,bennett1979dissipation,gaspard2016kinetics,gaspard2016kinetics1,gaspard2016kinetics2,Gaspard2009Molecular,Rao_2015,banerjee2017elucidating,chiuchiu2019error} }
\newcommand{\citeAS}{\cite{poulton2019nonequilibrium,poulton2021edge,ouldridge2017fundamental}}
\newcommand{\citePrevTech}{\cite{murugan2014discriminatory,murugan2012speed,yu2021energy,sahoo2013backtracking,sahoo2021accuracy,ehrenberg1980thermodynamic,song2016proofreading,song2020thermodynamic, li2019template,pigolotti2016protocols,sartori2015thermodynamics,wong2018energy,pineros2020kinetic,poulton2019nonequilibrium,bennett1979dissipation,wall1941structure,wall1944structure,mayo1944copolymerization,nguyen2016design,galstyan2019allostery,Rao_2015,banerjee2017elucidating,chiuchiu2019error} }
\newcommand{\citeSolvable}{\cite{murugan2014discriminatory,murugan2012speed,yu2021energy,sahoo2013backtracking,sahoo2021accuracy,ehrenberg1980thermodynamic,song2020thermodynamic, li2019template,pigolotti2016protocols,sartori2015thermodynamics,wong2018energy,pineros2020kinetic,poulton2019nonequilibrium,bennett1979dissipation,wall1941structure,wall1944structure,mayo1944copolymerization,nguyen2016design,galstyan2019allostery,mallory2020we,Rao_2015,banerjee2017elucidating,chiuchiu2019error} }
\newcommand{\citeTip}{\cite{poulton2019nonequilibrium,whitelam2012self,wall1941structure,wall1944structure,mayo1944copolymerization,nguyen2016design,Gaspard2009Molecular,gaspard2014kinetics,gaspard2016kinetics,gaspard2016kinetics1,gaspard2016kinetics2,gaspard2016template}}
\newcommand{\citeNHop}{\cite{yu2021energy,ehrenberg1980thermodynamic} }

\newcommand{\citeTemp}{\cite{sahoo2021accuracy,song2016proofreading,song2020thermodynamic,li2019template,pigolotti2016protocols,sartori2015thermodynamics,wong2018energy,pineros2020kinetic,bennett1979dissipation,gaspard2016kinetics,gaspard2016kinetics1,gaspard2016kinetics2,Gaspard2009Molecular,poulton2019nonequilibrium,poulton2021edge}}

\newcommand{\figtwoa}{\begin{tikzpicture}
    \tikzstyle{vertex}=[circle,draw=black, thick,inner sep=0pt,minimum size=12pt]
    \tikzstyle{edge}=[->, thick]
    \node[vertex] (1) at (0,0) {1};
    \node[vertex] (3) at (3,0) {3};
    \node[vertex] (2) at (1.5,2.598) {2};
    \node[vertex] (4) at (4.5,2.598) {4};
    \node[vertex] (B) at (5.5,2.598) {B};
    \node[vertex] (A) at (5.5,0) {A};
    \draw[transform canvas={yshift=0.25ex,xshift=-0.433ex},edge] (1)--node[above left]{$r_{12}$}(2);
    \draw[transform canvas={yshift=0.5ex},edge](1)--node[above]{$r_{13}$}(3);
    \draw[transform canvas={yshift=0.25ex,xshift=0.433ex},edge] (2)--node[above right]{$r_{23}$}(3);
    \draw[transform canvas={yshift=0.5ex},edge](2)--node[above]{$r_{24}$}(4);
    \draw[transform canvas={yshift=0.25ex,xshift=-0.433ex},edge](3)--node[above left]{$r_{34}$}(4);
    \draw[transform canvas={yshift=-0.25ex,xshift=0.433ex},edge] (2)--node[below right]{$r_{21}$}(1);
    \draw[transform canvas={yshift=-0.5ex},edge](3)--node[below]{$r_{31}$}(1);
    \draw[transform canvas={yshift=-0.25ex,xshift=-0.433ex},edge] (3)--node[below left]{$r_{32}$}(2);
    \draw[transform canvas={yshift=-0.5ex},edge](4)--node[below]{$r_{42}$}(2);
    \draw[transform canvas={yshift=-0.25ex,xshift=0.433ex},edge](4)--node[below right]{$r_{43}$}(3);
    \draw[edge] (4)--node[above]{$k_B$}(B);
    \draw[edge] (3)--node[above]{$k_A$}(A);
    \end{tikzpicture}}

\newcommand{\figtwob}{\begin{tikzpicture}
    \tikzstyle{vertex}=[circle,draw=black, thick,inner sep=0pt,minimum size=12pt]
    \tikzstyle{edge}=[->, thick]
    \node[vertex] (1) at (0,0) {1};
    \node[vertex] (3) at (3,0) {3};
    \node[vertex] (2) at (1.5,2.598) {2};
    \node[vertex] (4) at (4.5,2.598) {4};
    \draw[transform canvas={yshift=0.25ex,xshift=-0.433ex},edge] (1)--node[above left]{$r_{12}$}(2);
    \draw[transform canvas={yshift=0.5ex},edge](1)--node[above]{$r_{13}$}(3);
    \draw[transform canvas={yshift=0.25ex,xshift=0.433ex},edge] (2)--node[above right]{$r_{23}$}(3);
    \draw[transform canvas={yshift=0.5ex},edge](2)--node[above]{$r_{24}$}(4);
    \draw[transform canvas={yshift=0.25ex,xshift=-0.433ex},edge](3)--node[above left]{$r_{34}$}(4);
    \draw[transform canvas={yshift=-0.25ex,xshift=0.433ex},edge] (2)--node[below right]{$r_{21}$}(1);
    \draw[transform canvas={yshift=-0.5ex},edge, color=red](3)--node[below]{$\color{black}r_{31}\color{red}+k_A$}(1);
    \draw[transform canvas={yshift=-0.25ex,xshift=-0.433ex},edge] (3)--node[below left]{$r_{32}$}(2);
    \draw[transform canvas={yshift=-0.5ex},edge](4)--node[below]{$r_{42}$}(2);
    \draw[transform canvas={yshift=-0.25ex,xshift=0.433ex},edge](4)--node[below right]{$r_{43}$}(3);
    \draw[edge,color=red] (4) edge[out=-45,in=-45, min distance = 100pt] node[below right]{$k_B$}(1);
    \end{tikzpicture}}
    
\newcommand{\smallarrow}[1][3pt]{\mathrel{%
   \hbox{\rule[\dimexpr\fontdimen22\textfont2-.2pt\relax]{#1}{.4pt}}%
   \mkern-4mu\hbox{\usefont{U}{lasy}{m}{n}\symbol{41}}}}

\newcommand{\figtwoc}{\begin{tikzpicture}
    \tikzstyle{vertex}=[circle,draw=black, thick,inner sep=0pt,minimum size=12pt]
    \tikzstyle{edge}=[->, thick]
    \node[vertex] (c) at (0,0) {$\;\;1\smallarrow2\smallarrow3\smallarrow1\;\;$};
    \node[vertex] (4) at (3,1.732) {4};
    \draw[transform canvas={yshift=0.354ex,xshift=-0.354ex},edge](c)--node[above left]{$r_{24}+r_{34}$}(4);
    \draw[transform canvas={yshift=-0.354ex,xshift=0.354ex},edge](4)--node[below right]{$r_{42}+r_{43}+k_B$}(c);
    \end{tikzpicture}}
    
\newcommand{\figthreea}{\begin{tabular}{|ccc|}
    \hline
    \tikz[baseline=5ex]{
    \tikzstyle{vertex}=[circle,fill=black,inner sep=0pt,minimum size=4pt]
    \tikzstyle{edge}=[->, thick]
    \node[vertex, label=1] (1) at (0,0) {};
    \node[vertex, label=3] (3) at (6ex,0) {};
    \node[vertex, label=2] (2) at (3ex,5.196ex) {};
    \node[vertex, label=4] (4) at (9ex,5.196ex) {}; 
    \draw[edge](1)--(2);
    \draw[edge](2)--(4);
    \draw[edge](4)--(3); 
    }
    & &\\
    \tikz[baseline=5ex]{
    \tikzstyle{vertex}=[circle,fill=black,inner sep=0pt,minimum size=4pt]
    \tikzstyle{edge}=[->, thick]
    \node[vertex] (1) at (0,0) {};
    \node[vertex] (3) at (6ex,0) {};
    \node[vertex] (2) at (3ex,5.196ex) {};
    \node[vertex] (4) at (9ex,5.196ex) {}; 
    \draw[edge](1)--(2);
    \draw[edge](2)--(3);
    \draw[edge](4)--(3);    
    }
    &  
    \tikz[baseline=5ex]{
    \tikzstyle{vertex}=[circle,fill=black,inner sep=0pt,minimum size=4pt]
    \tikzstyle{edge}=[->, thick]
    \node[vertex] (1) at (0,0) {};
    \node[vertex] (3) at (6ex,0) {};
    \node[vertex] (2) at (3ex,5.196ex) {};
    \node[vertex] (4) at (9ex,5.196ex) {};
    \draw[edge](1)--(2);
    \draw[edge](2)--(3);
    \draw[edge] (4) edge[out=-45,in=-45, min distance = 30pt](1);  
    }
    &
    \tikz[baseline=5ex]{
    \tikzstyle{vertex}=[circle,fill=black,inner sep=0pt,minimum size=4pt]
    \tikzstyle{edge}=[->, thick]
    \node[vertex] (1) at (0,0) {};
    \node[vertex] (3) at (6ex,0) {};
    \node[vertex] (2) at (3ex,5.196ex) {};
    \node[vertex] (4) at (9ex,5.196ex) {}; 
    \draw[edge](1)--(2);
    \draw[edge](2)--(3);
    \draw[edge](4)--(2); 
    }
    \\
    \tikz[baseline=5ex]{
    \tikzstyle{vertex}=[circle,fill=black,inner sep=0pt,minimum size=4pt]
    \tikzstyle{edge}=[->, thick]
    \node[vertex] (1) at (0,0) {};
    \node[vertex] (3) at (6ex,0) {};
    \node[vertex] (2) at (3ex,5.196ex) {};
    \node[vertex] (4) at (9ex,5.196ex) {}; 
    \draw[edge](1)--(3);
    \draw[edge](2)--(1);
    \draw[edge](4)--(3);
    }
    &  
    \tikz[baseline=5ex]{
    \tikzstyle{vertex}=[circle,fill=black,inner sep=0pt,minimum size=4pt]
    \tikzstyle{edge}=[->, thick]
    \node[vertex] (1) at (0,0) {};
    \node[vertex] (3) at (6ex,0) {};
    \node[vertex] (2) at (3ex,5.196ex) {};
    \node[vertex] (4) at (9ex,5.196ex) {}; 
    \draw[edge](1)--(3);
    \draw[edge](2)--(1);
    \draw[edge] (4) edge[out=-45,in=-45, min distance = 30pt](1);
    }
    &
    \tikz[baseline=5ex]{
    \tikzstyle{vertex}=[circle,fill=black,inner sep=0pt,minimum size=4pt]
    \tikzstyle{edge}=[->, thick]
    \node[vertex] (1) at (0,0) {};
    \node[vertex] (3) at (6ex,0) {};
    \node[vertex] (2) at (3ex,5.196ex) {};
    \node[vertex] (4) at (9ex,5.196ex) {}; 
    \draw[edge](4)--(2);
    \draw[edge](2)--(1);
    \draw[edge](1)--(3);
    }
    \\
    \tikz[baseline=5ex]{
    \tikzstyle{vertex}=[circle,fill=black,inner sep=0pt,minimum size=4pt]
    \tikzstyle{edge}=[->, thick]
    \node[vertex] (1) at (0,0) {};
    \node[vertex] (3) at (6ex,0) {};
    \node[vertex] (2) at (3ex,5.196ex) {};
    \node[vertex] (4) at (9ex,5.196ex) {}; 
    \draw[edge](1)--(3);
    \draw[edge](2)--(3);
    \draw[edge](4)--(3);    
    }
    &  
    \tikz[baseline=5ex]{
    \tikzstyle{vertex}=[circle,fill=black,inner sep=0pt,minimum size=4pt]
    \tikzstyle{edge}=[->, thick]
    \node[vertex] (1) at (0,0) {};
    \node[vertex] (3) at (6ex,0) {};
    \node[vertex] (2) at (3ex,5.196ex) {};
    \node[vertex] (4) at (9ex,5.196ex) {};
    \draw[edge](1)--(3);
    \draw[edge](2)--(3);
    \draw[edge] (4) edge[out=-45,in=-45, min distance = 30pt](1);  
    }
    &
    \tikz[baseline=5ex]{
    \tikzstyle{vertex}=[circle,fill=black,inner sep=0pt,minimum size=4pt]
    \tikzstyle{edge}=[->, thick]
    \node[vertex] (1) at (0,0) {};
    \node[vertex] (3) at (6ex,0) {};
    \node[vertex] (2) at (3ex,5.196ex) {};
    \node[vertex] (4) at (9ex,5.196ex) {}; 
    \draw[edge](4)--(2);
    \draw[edge](2)--(3);
    \draw[edge](1)--(3); 
    }
    \\
    \tikz[baseline=5ex]{
    \tikzstyle{vertex}=[circle,fill=black,inner sep=0pt,minimum size=4pt]
    \tikzstyle{edge}=[->, thick]
    \node[vertex] (1) at (0,0) {};
    \node[vertex] (3) at (6ex,0) {};
    \node[vertex] (2) at (3ex,5.196ex) {};
    \node[vertex] (4) at (9ex,5.196ex) {}; 
    \draw[edge](2)--(4);
    \draw[edge](4)--(3);
    \draw[edge](1)--(3); 
    }
    &  
    \tikz[baseline=5ex]{
    \tikzstyle{vertex}=[circle,fill=black,inner sep=0pt,minimum size=4pt]
    \tikzstyle{edge}=[->, thick]
    \node[vertex] (1) at (0,0) {};
    \node[vertex] (3) at (6ex,0) {};
    \node[vertex] (2) at (3ex,5.196ex) {};
    \node[vertex] (4) at (9ex,5.196ex) {};
    \draw[edge](2)--(4);
    \draw[edge] (4) edge[out=-45,in=-45, min distance = 30pt](1);
    \draw[edge](1)--(3);
    }
    &
    \\\hline
\end{tabular} }

\newcommand{\figthreeb}{\begin{tabular}{|ccc|}
    \hline
    \tikz[baseline=5ex]{
    \tikzstyle{vertex}=[circle,fill=black,inner sep=0pt,minimum size=4pt]
    \tikzstyle{edge}=[->, thick]
    \node[vertex] (1) at (0,0) {};
    \node[vertex] (3) at (6ex,0) {};
    \node[vertex] (2) at (3ex,5.196ex) {};
    \node[vertex] (4) at (9ex,5.196ex) {}; 
    \draw[edge](1)--(2);
    \draw[edge](2)--(3);
    \draw[edge](3)--(4);  
    }
    &&
    \\&&\\
    \tikz[baseline=5ex]{
    \tikzstyle{vertex}=[circle,fill=black,inner sep=0pt,minimum size=4pt]
    \tikzstyle{edge}=[->, thick]
    \node[vertex] (1) at (0,0) {};
    \node[vertex] (3) at (6ex,0) {};
    \node[vertex] (2) at (3ex,5.196ex) {};
    \node[vertex] (4) at (9ex,5.196ex) {}; 
    \draw[edge](3)--(1);
    \draw[edge](1)--(2);
    \draw[edge](2)--(4);
    }
    &  
    \tikz[baseline=5ex]{
    \tikzstyle{vertex}=[circle,fill=black,inner sep=0pt,minimum size=4pt]
    \tikzstyle{edge}=[->, thick]
    \node[vertex] (1) at (0,0) {};
    \node[vertex] (3) at (6ex,0) {};
    \node[vertex] (2) at (3ex,5.196ex) {};
    \node[vertex] (4) at (9ex,5.196ex) {}; 
    \draw[edge](3)--(2);
    \draw[edge](1)--(2);
    \draw[edge](2)--(4);
    }
    &
    \tikz[baseline=5ex]{
    \tikzstyle{vertex}=[circle,fill=black,inner sep=0pt,minimum size=4pt]
    \tikzstyle{edge}=[->, thick]
    \node[vertex] (1) at (0,0) {};
    \node[vertex] (3) at (6ex,0) {};
    \node[vertex] (2) at (3ex,5.196ex) {};
    \node[vertex] (4) at (9ex,5.196ex) {}; 
    \draw[edge](3)--(4);
    \draw[edge](1)--(2);
    \draw[edge](2)--(4);
    }
    \\&&\\
    \tikz[baseline=5ex]{
    \tikzstyle{vertex}=[circle,fill=black,inner sep=0pt,minimum size=4pt]
    \tikzstyle{edge}=[->, thick]
    \node[vertex] (1) at (0,0) {};
    \node[vertex] (3) at (6ex,0) {};
    \node[vertex] (2) at (3ex,5.196ex) {};
    \node[vertex] (4) at (9ex,5.196ex) {}; 
    \draw[edge](2)--(1);
    \draw[edge](1)--(3);
    \draw[edge](3)--(4);    
    }
    &  
    \tikz[baseline=5ex]{
    \tikzstyle{vertex}=[circle,fill=black,inner sep=0pt,minimum size=4pt]
    \tikzstyle{edge}=[->, thick]
    \node[vertex] (1) at (0,0) {};
    \node[vertex] (3) at (6ex,0) {};
    \node[vertex] (2) at (3ex,5.196ex) {};
    \node[vertex] (4) at (9ex,5.196ex) {};
    \draw[edge](2)--(3);
    \draw[edge](1)--(3);
    \draw[edge](3)--(4);  
    }
    &
    \tikz[baseline=5ex]{
    \tikzstyle{vertex}=[circle,fill=black,inner sep=0pt,minimum size=4pt]
    \tikzstyle{edge}=[->, thick]
    \node[vertex] (1) at (0,0) {};
    \node[vertex] (3) at (6ex,0) {};
    \node[vertex] (2) at (3ex,5.196ex) {};
    \node[vertex] (4) at (9ex,5.196ex) {}; 
    \draw[edge](2)--(4);
    \draw[edge](1)--(3);
    \draw[edge](3)--(4);  
    }
    \\&&\\
    \tikz[baseline=5ex]{
    \tikzstyle{vertex}=[circle,fill=black,inner sep=0pt,minimum size=4pt]
    \tikzstyle{edge}=[->, thick]
    \node[vertex] (1) at (0,0) {};
    \node[vertex] (3) at (6ex,0) {};
    \node[vertex] (2) at (3ex,5.196ex) {};
    \node[vertex] (4) at (9ex,5.196ex) {};
    \draw[edge](1)--(3);
    \draw[edge](3)--(2);
    \draw[edge](2)--(4);
    }
    &&
    \\&&\\\hline
\end{tabular} }

\usepackage{caption}
\usepackage{subcaption}
\usepackage{nicefrac}
\usepackage{mathtools}

\makeatletter
\def\@email#1#2{%
 \endgroup
 \patchcmd{\titleblock@produce}
  {\frontmatter@RRAPformat}
  {\frontmatter@RRAPformat{\produce@RRAP{*#1\href{mailto:#2}{#2}}}\frontmatter@RRAPformat}
  {}{}
}%
\makeatother
\begin{document}

\preprint{AIP/123-QED}

\title[A Universal Method for Analysing Copolymer Growth]{A Universal Method for Analysing Copolymer Growth}

\author{Benjamin Qureshi}
\affiliation{Department of Bioengineering and Centre for Synthetic Biology, Imperial College London, London SW7 2AZ, United Kingdom}
\author{Jordan Juritz}
\affiliation{Department of Bioengineering and Centre for Synthetic Biology, Imperial College London, London SW7 2AZ, United Kingdom}
\author{Jenny M. Poulton}
\affiliation{Foundation for Fundamental Research on Matter (FOM) Institute for Atomic and Molecular Physics (AMOLF), 1098 XE Amsterdam, The Netherlands}
\author{Adrian Beersing-Vasquez}
\affiliation{Faculty of Science, Amsterdam Science Park 904, 1098 XH Amsterdam}

\author{Thomas E. Ouldridge}%
\email{t.ouldridge@imperial.ac.uk}
\affiliation{Department of Bioengineering and Centre for Synthetic Biology, Imperial College London, London SW7 2AZ, United Kingdom}

\date{\today}

\begin{abstract}
Polymers consisting of more than one type of monomer, known as copolymers, are vital to both living and synthetic systems. Copolymerisation has been studied theoretically in a number of contexts, often by considering a Markov process in which monomers are added or removed from the growing tip of a long copolymer. To date, the analysis of the most general models of this class has necessitated simulation. We present a general method for analysing such processes without resorting to simulation. Our method can be applied to models with an arbitrary network of sub-steps prior to addition or removal of a monomer, including non-equilibrium kinetic proofreading cycles. Moreover,  the approach allows for a dependency of addition and removal reactions on the neighbouring site in the copolymer, and thermodynamically self-consistent models in which all steps are assumed to be microscopically reversible. Using our approach, thermodynamic quantities such as chemical work; kinetic quantities such as time taken to grow; and statistical quantities such as the distribution of monomer types in the growing copolymer can be derived either analytically or numerically directly from the model definition.
\end{abstract}

\maketitle

\section{Introduction}

\begin{figure*}
    \centering
    \includegraphics[scale=0.60]{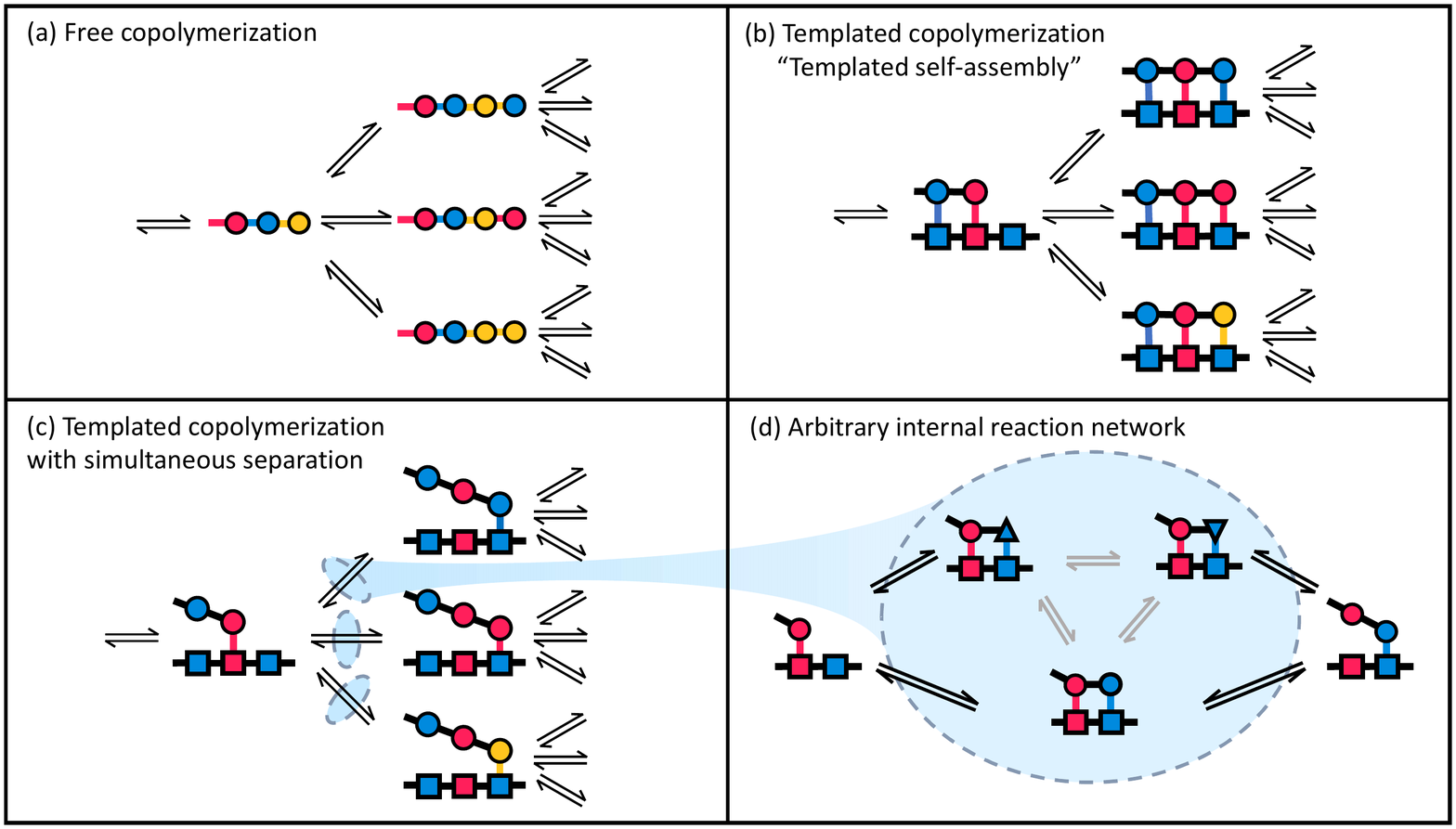}
    \caption{ Comparison of the different types of copolymerisation mechanism with three types of monomer (blue, red and yellow). (a) shows free copolymerisation, (b) templated self assembly, and (c) templated copolymerisation with autonomous separation. In (b) and (c) the template is shown with squares and the growing polymer with circles. (d) An example of a more detailed reaction scheme used to select the next monomer. In each of these sub-figures, different colours represent different monomer types, with bonds coloured accordingly when their strength might depend on the monomer. In (d), different activation states of the monomer undergoing incorporation are represented by different shapes. The dashed bubble indicates how the arbitrary set of reactions in (d) may replace the simple reaction surrounded by a dashed bubble in (c) for a more complex model.}
    \label{fig:modeltypes}
\end{figure*}

Copolymers are polymers consisting of more than one type of monomeric unit; the order of these monomers in the chain defines the copolymer {\it sequence}. Broadly, copolymerisation mechanisms can be classified into two main categories: free copolymerisation that does not rely on a template\cite{gaspard2016kinetics}, as shown in figure~\ref{fig:modeltypes}(a); and templated copolymerisation, in which a template (usually another copolymer) is used to bias the distribution of sequences produced, as shown in figure~\ref{fig:modeltypes}(b) and figure~\ref{fig:modeltypes}(c). Polymers produced via both types of mechanism are of relevance to both biological and industrial systems. In living systems, O-glycans are sequences of monosaccharides that grow by free copolymerisation from serine or threonine amino acids\cite{CORFIELD2015236}. They play a key role as a physical protective barrier for cells from pathogens, as well as participating in other cellular processes\cite{CORFIELD2015236,MucinsReview}. Free copolymerisation is also a common method for producing plastics and rubbers in commercial and industrial systems\cite{chanda2013introduction,overberger1985copolymerization}. Additionally, there have been recent experimental designs for free copolymerisation systems to produce specific products utilising DNA-nanotechnology-based reaction schemes\cite{meng2016autonomous,zhang2019programming}. 

Templated copolymerisation is the mechanism by which DNA, RNA and polypeptides are produced in DNA replication, RNA transcription, and protein translation, respectively. These processes are at the heart of the central dogma of molecular biology\cite{crick1970central} and are the basis of the informational and biochemical complexity of life. In DNA replication, DNA templates the production of copies of itself; in transcription, DNA templates the production of RNA; and, in translation, mRNA is the template for the production of a polypeptide\cite{MolBiolCell}. Inspired by these biological templated copolymerisation mechanisms, there has been recent interest in designing synthetic systems that can produce other sequence-controlled molecules via templated copolymerisation\cite{niu2013enzyme,kong2016generation, stross2017sequence, lutz2018sequence, nunez2019sequence, nunez2021replication,cabello2021handhold}.
 
Free polymerisation can be modelled as a Markovian growth process under which monomers bind to the end of a growing polymer at a certain rate. Early free copolymerisation models\cite{wall1941structure,wall1944structure,mayo1944copolymerization} built on this framework to allow for copolymerisation via the incorporation of multiple types of monomeric unit, as shown in figure~\ref{fig:modeltypes}(a), albeit with irreversible polymerisation reactions. In particular, Mayo and Lewis\cite{mayo1944copolymerization} emphasised that in polymerisation models, if the monomer binding events are irreversible and their rates are conditional on the terminal monomer type, then intra-sequence correlations are generated within the copolymer.

Although the use of models with irreversible transitions is reasonable in many contexts,
thermodynamically self-consistent models require all transitions to be microscopically reversible.\cite{ouldridge2018importance} Specifically, if a transition from state $A$ to state $B$ is possible, then transitions from $B$ to $A$ must also be possible. Models with fully microscopically reversible polymerisation reactions, as in figure~\ref{fig:modeltypes}(a),  are more challenging to analyse but can be interpreted in a thermodynamic sense.\cite{whitelam2012self,gaspard2016kinetics,nguyen2016design}

Templates can affect the rate at which monomers are added or removed from a growing copolymer, and hence templated copolymerisation models can be more complex than free copolymerisation models. When the template consists of just one type of templating monomer (homopolymer), a templated copolymerisation process can be mapped onto a free copolymerisation model. Further, if one assumes some symmetries regarding interactions between monomers in the growing copolymer and those in the template (such as all complementary bonds have equal strength and all non-complementary bonds have equal strength),  models of sequence-bearing templates may be mapped onto models with homopolymeric templates, and hence to models of free copolymerisation\cite{poulton2019nonequilibrium,poulton2021edge,juritz2022,sartori2013kinetic,sartori2015thermodynamics}. 
 
Templated copolymerisation models can be further divided into two main categories: templated self-assembly (figure~\ref{fig:modeltypes}(b)\citeTSA and autonomously-separating mechanisms (figure~\ref{fig:modeltypes}(c)\citeAS. Templated self-assembly models are those in which all the monomers in the growing copolymer remain bound to the template. In autonomously-separating models, the growing copolymer detaches as it extends \cite{poulton2019nonequilibrium,poulton2021edge,juritz2022}. There has been recent interest in explicitly modelling autonomous separation in templated growth in an attempt to understand models that give a better description of transcription or translation\cite{poulton2019nonequilibrium,poulton2021edge,juritz2022}. In autonomously-separating models, the simultaneous growth and separation of the copolymer and template mean that the copy-template interactions are not permanent, and therefore free energy released from such interactions cannot be part of the driving force of polymerisation. Additionally, since these copy-template bonds are temporary, they cannot stabilise the accurate copy directly in the long time limit. Further, an ensemble of accurate polymers is a lower entropy state than an ensemble of random polymers. These conditions mean that non-equilibrium driving is required to generate accurate copies of the template if the copies are to spontaneously detach\cite{ouldridge2017fundamental}. Moreover, the separation of the lagging tail from the template as the copolymer grows naturally causes intra-sequence correlations within the product.\cite{poulton2019nonequilibrium} 
 
The models described above are maximally coarse-grained, in that they treat the binding of monomers to the growing tip of the copolymer as a simple, usually single-step, process. However, more generally, one may wish to study models in which polymerisation occurs via a more detailed series of steps, as in figure~\ref{fig:modeltypes}(d). For instance, in order to explain the high accuracy observed in biological polymer copying systems, Hopfield\cite{hopfield1974kinetic} and Ninio\cite{ninio1975kinetic} independently introduced the concept of kinetic proofreading: a reaction motif in which a monomer undergoes a free energy consuming activation reaction before it is polymerised into the copolymer. The introduction of kinetic proofreading reaction motifs presaged the investigation of more complex copolymerisation mechanisms\cite{bennett1979dissipation,mallory2020we}.

In summary, models that allow for multiple monomer types, intra-sequence correlations, reversible reactions, and general, multi-step monomer inclusion reactions represent a wide class of copolymerisation processes. Previous  techniques \citePrevTech 
 have not allowed analysis of thermodynamically self-consistent models of generalised free copolymerisation processes in which monomer addition is given by an arbitrarily complex network of reversible reactions with rates that may depend on the terminal monomer type, and templated copolymerisation models with high symmetry that can be mapped to these free processes. Investigating the most general type of model in this class would require simulation.
 
In this paper we present a universal method for studying this large class of copolymerisation models. Drawing on the work of Gaspard and Andrieux\cite{gaspard2014kinetics} for analysing linear copolymerisation processes, and Hill\cite{hill1966studies,hill1988interrelations} for analysing absorbing Markov processes, we present analytical methods for extracting: explicit expressions for the probability of inclusion of a given monomer; the growth rate of a copolymerisation process; and the chemical work done by the process. Our method removes the need to extract the same features by simulation and often produces simple, analytic results. 

In section~\ref{Absorbing}, we review and refine methods relating to absorbing Markov chains that are crucial to understanding our approach. In section~\ref{method} we present our method.  In section~\ref{examples}, we apply the method to a few example processes to demonstrate its use and power when considering models with certain features. First, we apply the method to models for which the rate of adding new monomers only depends on the monomer type being added. Next we apply the method to templated copolymerisation systems with autonomous-separation that do not have non-equilibrium kinetic proofreading cycles. Finally, we solve a generalised version of Hopfield's kinetic proofreading model applied to a templated copolymerisation system with an autonomously separating product.

\section{Methods}
\begin{figure}
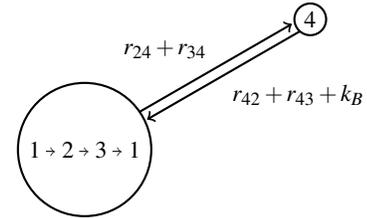

    \centering
    \begin{subfigure}[b]{0.3\textwidth}
    \caption{Example absorbing Markov process}
    
    \figtwoa
    \end{subfigure}
    
    \begin{subfigure}[b]{0.3\textwidth}
    \vspace{10pt}
    \caption{Closed process}
    \figtwob
    \end{subfigure}
    
    \begin{subfigure}[b]{0.3\textwidth}
    \vspace{-30pt}
    \caption{Cycle process for cycle $1\to2\to3\to1$}
    \figtwoc
    \end{subfigure}
    \caption{Graphical representations of an absorbing Markov process to illustrate the methodology outlined in section~\ref{Absorbing}. (a) Example absorbing Markov process $(\mathcal{X}\cup\mathcal{A},K)$, with two absorbing states, $\mathcal{A}=\{A,B\}$, and four transient states $\mathcal{X}=\{1,2,3,4\}$. (b) The closed process starting at state 1, $(\mathcal{X},K_1)$. (c) The cycle process $(\{c\}\cup\mathcal{X}/\{1,2,3\}=\{c,4\}, K_{1,C})$ for the cycle $C=1\smallarrow2\smallarrow3\smallarrow1$ or $C'=1\smallarrow3\smallarrow2\smallarrow1$. }
    \label{fig:example}
    
\end{figure}
\subsection{Absorbing Markov Chains}
\label{Absorbing}

We begin by reviewing and adapting some diagrammatic techniques introduced by Hill to analyse absorbing Markov chains\cite{hill1966studies,hill1988interrelations}. An absorbing Markov chain is a Markov chain for which any trajectory through its state space with arbitrary initial conditions will reach an absorbing state in finite time almost surely\cite{kemeny1983finite}. We can decompose the state space of an absorbing Markov chain into absorbing states, $\mathcal{A},$ and transient states, $\mathcal{X}$, such that the state space is $V=\mathcal{A}\cup\mathcal{X}$. Let us denote the rate function that describes the chain as $K:V\times V \to \mathbb{R}^+$, such that $K(x,y)$ is the rate of the transition from state $x$ to state $y$. Then we denote a Markov process as the tuple, $(V,K)$.
 
Throughout this section, we shall refer to the absorbing Markov chain given in figure~\ref{fig:example}(a), which possesses two absorbing states and non-trivial cycles, for illustrative purposes.
 
\subsubsection{Expectations of an absorbing process are steady-state averages of a ``closed process"}
We will derive expressions for four main quantities: the probabilities of reaching certain absorbing states, the expected time taken to absorption, the expected net number of times traversing a given edge before absorption and the expected number of times that a trajectory goes round a cycle before absorption. These quantities depend on the starting (transient) state $\sigma\in\mathcal{X}$ and can be found in terms of the "closed" process\cite{hill1988interrelations}. The closed process is a modified version of an absorbing Markov chain in which transitions to the absorbing states are redirected to the starting state. Figure~\ref{fig:example}(b) shows the closed process starting at state 1 of our example absorbing chain of figure~\ref{fig:example}(a). The closed process for a Markov process $(\mathcal{X}\cup\mathcal{A},K)$ starting at state $\sigma$ is a new Markov process $(\mathcal{X},K_\sigma)$ with a rate function given by:
\begin{eqnarray}
    K_\sigma(x,\sigma)=K(x,\sigma)+\sum_{A\in\mathcal{A}}K(x,A)
    \label{eq:ClosedRate}
\end{eqnarray}
for $x\in\mathcal{X}$ and agreeing with $K$ on $\mathcal{X}\times\mathcal{X}\slash\{\sigma\}$. 
 
The closed process has a unique stationary distribution for the following reasons. From the definition of an absorbing Markov chain, there exists a path from any state to an absorbing state, taking finite time. Thus, in the closed process, there is a path from any state to the starting state, taking finite time. The set of states including the starting state and all those that may be reached from the starting state is therefore positive recurrent and further, this set is the only recurrent set of states and will be reached from any other state. Since there is only one recurrent set of states, there is a unique stationary distribution\cite{kemeny1983finite}. 

\begin{figure*}
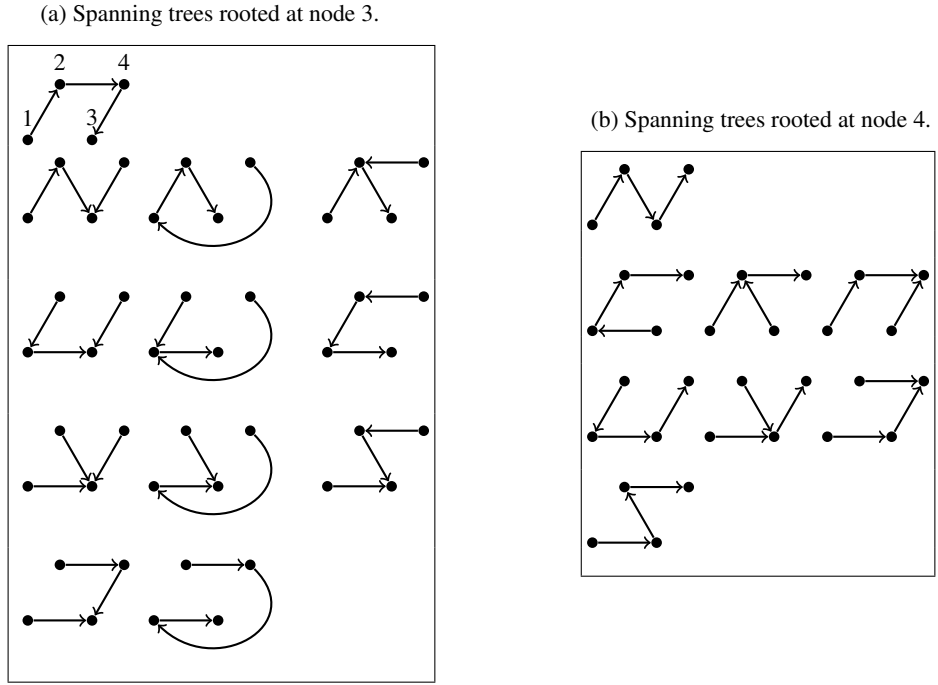

    \centering
    \begin{subfigure}[b]{0.3\textwidth}
    \caption{Spanning trees rooted at node 3.}
    \label{fig:SpanTreeA}
    \figthreea
    \end{subfigure}
    \hspace{0.1\textwidth}
    \begin{subfigure}[b]{0.3\textwidth}
    \caption{Spanning trees rooted at node 4.}
    \figthreeb
    \label{fig:SpanTreeB}
    \end{subfigure}
    \caption{Spanning trees of the closed processes rooted at (a) node 3 and (b) node 4 derived from figure~\ref{fig:example}\,(b), with nodes labelled in the first spanning tree and all other trees following the same positioning. The spanning trees have been arranged in terms of the self-avoiding walk between nodes 1 and 3 for the trees rooted at node 3 and arranged in terms of the self-avoiding walks between nodes 1 and 4 for the trees rooted at node 4. More details on the relationship between self avoiding walks and spanning trees are given in appendix~\ref{App:SAW}.}
    \label{fig:SpanTree}
\end{figure*}

Expected quantities of an absorbing Markov chain, such as the expected probability that a particular absorbing state is reached,  can be found in terms of steady state quantities in the closed process. Whenever a trajectory of the original process reaches an absorbing state, in the closed process that same trajectory would have been reset back to the starting state. Hence, running the closed process for long times is equivalent to generating many independent trajectories to absorption for the original chain. Thus, averaging quantities in the steady state of the closed process is equivalent to taking expectations over independent trials of quantities in the absorbing chain. It is worth noting that the dependence of expected quantities on the starting state is encoded in the definition of the closed process. Finally, we can see that the definition of the closed process may permit self-transitions, $\sigma\to\sigma$, which, for continuous time Markov processes, have little meaning. However, for the purposes of calculating steady state probabilities of the closed process they may be ignored. 
 
\subsubsection{Steady State averages of the closed process are calculated using the Markov chain tree theorem}
Given that we can turn the calculation of expectations of absorbing processes into steady-state averages over closed processes, we can make use of tools developed for analysing the steady state of Markov processes, such as the Markov chain tree theorem (MCTT)\cite{anantharam1989proof}. The MCTT states that the steady state distribution of a Markov chain with a unique stationary distribution may be found by summing over rooted spanning trees of the process, where the transition rates are taken as weights on the edges of the graph. Explicitly, let $\mathcal{G}$ be a directed weighted graph, with weight $K(e)$ for an edge $e$ of $\mathcal{G}$. A spanning tree of $\mathcal{G}$, rooted at a vertex, $\nu$, is a subgraph of $\mathcal{G}$ with no cycles that connects all the vertices of $\mathcal{G}$ and for which the out degree of every vertex, except $\nu$, is one. The sets of spanning trees rooted at nodes 3 and 4 of the closed process, figure~\ref{fig:example}(b), are shown in figure~\ref{fig:SpanTree}. Denote by $\mathcal{T}(x)$ the set of all spanning trees rooted at $x$. The MCTT states that the steady state probability to be in state $x$, $\pi(x)$, is given by:
\begin{eqnarray}
    \pi(x)=\frac{\sum\limits_{T\in\mathcal{T}(x)}\prod\limits_{e\in T}K(e)}{\sum\limits_{v\in\mathcal{X}}\sum\limits_{T\in\mathcal{T}(v)}\prod\limits_{e\in T}K(e)},
    \label{eqn:MCTTss}
\end{eqnarray}
with $e\in T$ representing the edges of the tree. The denominator here is simply a normalisation constant. 
 
We can define steady state currents from the steady state distribution of the closed process that corresponds to expected currents of the absorbing chain. Let a subscript $\sigma$ denote quantities in the closed process starting at state $\sigma$. Then $\pi_\sigma$ is the steady state probability distribution and $K_\sigma$ the rate function. The current along a given edge, $e=x\to y$, is given by the probability to be in state $x$, $\pi_\sigma(x)$, multiplied by the rate along said edge. We can, therefore, write the steady state current along all edges that originally led to absorbing states as:
\begin{eqnarray}
   J_{\text{Tot}}(\sigma) = \sum_{A\in\mathcal{A}}\sum_{x\in\mathcal{X}}\pi_\sigma(x)K(x,A),
   \label{eq:AbsorbingFlux}
\end{eqnarray}
where, as before, $\mathcal{X}$ is the set of transient states and $\mathcal{A}$, the set of absorbing states. $J_{\text{Tot}}(\sigma)$ is the expected total current to absorbing states from state $\sigma$, and, therefore, its reciprocal is the expected time to absorption.
  
For the example process shown in figure~\ref{fig:example}, we present the spanning trees of the corresponding closed process rooted at node 3 and 4 in figure~ \ref{fig:SpanTree}. Given the spanning trees, we can directly write down the total current to absorbing states as:
\begin{eqnarray}
    \nonumber
    J&_{\text{Tot}}&(1)=\frac{1}{\mathcal{N}}\Big[k_A\big[r_{12}r_{24}r_{43}+r_{12}r_{23}(r_{42}+r_{43}+k_B)\\\nonumber
    &+&r_{13}((r_{43}+k_B)(r_{21}+r_{23}+r_{24})+r_{42}(r_{21}+r_{23}))\big]\\\nonumber
    &+&k_B\big[r_{12}r_{23}r_{34}+r_{12}r_{24}(r_{31}+k_A+r_{32}+r_{34})\\
    &+&r_{13}r_{34}(r_{21}+r_{23}+r_{24})+r_{13}r_{32}r_{24}\big]\Big],
\end{eqnarray}
where $\mathcal{N}$ is the normalisation term, given in appendix~\ref{App:ExNorm}. The terms multiplied by $k_A$ are the partial current to absorbing state $A$, i.e. the current along transition $3\to A$, coming from the trees rooted at node $3$, and equivalently for $k_B$ with state $B$, i.e. the current along transition $4\to B$, coming from trees rooted at node $4$.
 
\subsubsection{Absorbing probabilities}
Given a Markov chain with multiple absorbing states, we can ask for the probability of absorption in each absorbing state in the long time limit. The probability that a trajectory eventually ends in a specific absorbing state  can be calculated from the closed process, by dividing the expected current along transitions that originally led to  the absorbing state in question by $J_{\text{Tot}}(\sigma)$ (eqn.~\ref{eq:AbsorbingFlux}). Therefore the absorption probabilities can be written
\begin{eqnarray}
    \mathbb{P}[\sigma\to A]=\frac{\sum\limits_{x\in\mathcal{X}}\pi_\sigma(x)K(x,A)}{\sum\limits_{B\in\mathcal{A}}\sum\limits_{x\in\mathcal{X}}\pi_\sigma(x)K(x,B)},
    \label{eq:absprob}
\end{eqnarray}
using the notation $\mathbb{P}[\sigma\to A]$ to denote probability of being absorbed to $A$ given that the trajectory started in state $\sigma$. It is worth noting here that given that this quantity is a ratio of currents, there is a factor of $\pi_\sigma$ in both the denominator and the numerator of the expression. In practice, we see that the normalisation factor from the MCTT (eqn.~\ref{eqn:MCTTss}) cancels out, which simplifies the quantities in the calculation.
 
For our example process shown in figure~\ref{fig:example}, we can use the partial currents to absorbing states $A$ and $B$ to write down the absorbing probabilities:
\begin{eqnarray}
\nonumber
    &\mathbb{P}&[1\to A]\;=\frac{1}{\mathcal{N}J_{\text{Tot}}(1)}k_A\big[r_{12}r_{24}r_{43}+r_{12}r_{23}(r_{42}+r_{43}+k_B)\\\nonumber
    &+&r_{13}((r_{43}+k_B)(r_{21}+r_{23}+r_{24})+r_{42}(r_{21}+r_{23}))\big],\\\nonumber
    &\mathbb{P}&[1\to B]\;=\frac{1}{\mathcal{N}J_{\text{Tot}}(1)}k_B\big[r_{13}r_{34}(r_{21}+r_{23}+r_{24})+r_{13}r_{32}r_{24}\\
    &+&r_{12}r_{23}r_{34}+r_{12}r_{24}(r_{31}+k_A+r_{32}+r_{34})\big]\Big].
\end{eqnarray}
The normalisation factor, $\mathcal{N}$, propagated through from eqn.~\ref{eqn:MCTTss}, conveniently cancels out with the $1/{\mathcal{N}}$ implicit in $J_{\text{tot}}$.
\vspace{-1mm}

\subsubsection{Counting edge and cycle transitions}
\label{sec:absorbingwork}
In this subsection, we shall calculate the expected net number of times traversing a given edge of an absorbing Markov process before absorption. Additionally, we shall calculate the expected number of times a non-recurrent cycle of an absorbing Markov process is traversed before absorption. Both of these will be of use later in defining a notion of chemical work.

To calculate the net number of times crossing a given edge we find the expected current along the transition, $x\leftrightharpoons y$, between states $x$ and $y$ of an absorbing process, $(\mathcal{X}\cup\mathcal{A},K)$, as in section~\ref{Absorbing}. The expected current through this edge, denoted $J_{x\leftrightharpoons y}(\sigma)$, given starting in state $\sigma\in\mathcal{X}$, can be calculated from the closed process, $(\mathcal{X},K_\sigma)$, as in eqn.~\ref{eq:ClosedRate}, as the difference between the steady state probability to be in state $x$ multiplied by the rate from $x\to y$ and the steady state probability to be in state $y$ multiplied by the rate from $y\to x$,
\begin{eqnarray}
        J_{x\leftrightharpoons y}(\sigma)=\pi_\sigma(x)K(x,y)-\pi_\sigma(y)K(y,x).
        \label{eq:xyflux}
\end{eqnarray}
The net number of times traversing this edge (number of observed transitions $x\to y$ - number of observed transitions $y\to x$) before absorption is then just the ratio between this current and total current to absorbing states:
\begin{eqnarray}
        N_{x\leftrightharpoons y}(\sigma)=\frac{ J_{x\leftrightharpoons y}(\sigma)}{J_{\text{Tot}}(\sigma)}.
        \label{eq:edgecount}
\end{eqnarray}
The current, eqn.~\ref{eq:xyflux}, is intimately linked to the notion of cycles as pointed out by Wachtel et al.\cite{Wachtel2018Thermodynamically} and detailed in appendix~\ref{App:Edge-Cycle}. Thus, we also wish to find the expected number of times traversing a non-recurrent cycle. We define a non-recurrent cycle for a Markov chain to be a cycle of states, where each state, aside from the originating state, does not appear more than once in the cycle. For example, the cycle A$\rightarrow$B$\rightarrow$C$\rightarrow$D$\rightarrow$A is non-recurrent, but A$\rightarrow$B$\rightarrow$C$\rightarrow$D$\rightarrow$B$\rightarrow$A is recurrent. Note that the originating state is arbitrary, and so A$\rightarrow$B$\rightarrow$C$\rightarrow$D$\rightarrow$A is equivalent to B$\rightarrow$C$\rightarrow$D$\rightarrow$A$\rightarrow$B. For a stationary process, the expected frequency with which a cycle is completed can be calculated from the one-way cycle current\cite{kohler1980frequency,hill1988interrelations}, which is the probability current going around the cycle. For a chosen non-recurrent cycle, the one-way cycle current can be calculated diagrammatically from three terms. First, a cycle term given by the product of rates around the cycle in the chosen direction. Second, a spanning tree term that can be found by collapsing the nodes in the cycle into a single node (in figure~\ref{fig:example}c, the cycle $1\to2\to3\to1$ has been collapsed in this way) and finding the sum of spanning trees of this new graph rooted at the collapsed cycle node. Finally, there is a normalisation factor, which is the same normalisation factor as for the current, $\mathcal{N}$. Explicitly, consider an absorbing Markov chain, $\mathcal{G}=(\mathcal{X}\cup\mathcal{A},K)$ and its closed process starting at $\sigma\in\mathcal{X}$, $\mathcal{G}_\sigma=(\mathcal{X},K_\sigma)$. Let $C$ denote both the set of edges and set of nodes of a cycle in the closed process. To calculate the spanning tree term for the one-way cycle current, construct a new Markov chain, the cycle process, $\mathcal{G}_{\sigma,C}=(\{c\}\cup(\mathcal{X}/C), K_{\sigma,C})$, where $\{c\}\cup(\mathcal{X}/C)$ is the set of transient states of the original process with the states in the cycle replaced by the single node, $c$, and $K_{\sigma,C}$ is given by:
\begin{eqnarray}
\nonumber
    K_{\sigma,C}(x,c)&=&\sum_{i\in C}K_\sigma(x,i)\\\nonumber
    K_{\sigma,C}(c,x)&=&\sum_{i\in C}K_\sigma(i,x)\\
    K_{\sigma,C}(c,c)&=&0
    \label{eq:cycleprocess}
\end{eqnarray}
for $x\in\mathcal{X}/C$ and agreeing with $K_\sigma$ elsewhere. The cycle process for the cycle $C=1231$ (or $C'=1321$) of the example system in figure~\ref{fig:example}(a) is shown in figure~\ref{fig:example}(c). Let $\mathcal{T}_\sigma(x),\mathcal{T}_C(x)$ be the sets of spanning trees rooted at $x$ of the closed process, $\mathcal{G}_\sigma$, and cycle process, $\mathcal{G}_{\sigma,C}$, respectively. Then, the cycle current is given by\cite{kohler1980frequency}:
\begin{eqnarray}
    J_{\text{Cyc}}(\sigma,C)=\frac{\overbrace{\left(\prod\limits_{e\in C}K(e)\right)}^{\text{Cycle}}\overbrace{\sum\limits_{T\in\mathcal{T}_C}\prod\limits_{e\in T}K_{\sigma,C}(e)}^{\text{Spanning Trees}}}{\underbrace{\sum\limits_{x\in\mathcal{X}}\sum\limits_{T\in\mathcal{T}_\sigma(x)}\prod\limits_{e\in T}K_{\sigma}(e)}_{\text{Normalisation}}}.
    \label{eqn:cycleflux}
\end{eqnarray}
Note for the cycle term, the edges are taken from the original process rather than the closed process. Given the cycle current for the closed process, the expected number of circulations of the cycle before absorption, $N_{\text{Cyc}}(\sigma,C)$, is the ratio of the cycle current to the total current to absorbing states:
\begin{eqnarray}
        N_{\text{Cyc}}(\sigma,C) = \frac{J_{\text{Cyc}}(\sigma,C)}{J_{\text{Tot}(\sigma)}}.
    \label{eqn:cyclecount}
\end{eqnarray} 
For our example process, the expected number of circulations of $C=1231$ is
\begin{eqnarray}
    N_{\text{Cyc}}(1,C)= \frac{(r_{12}r_{23}r_{31})(r_{42}+r_{43}+k_B)}{\mathcal{N}J_{\text{Tot}}(1)},
\end{eqnarray}
with the same implicit cancellation of normalisation as before, since $J_{\text{Tot}}\propto\frac{1}{\mathcal{N}}$.

For an absorbing process starting at a given state, we may divide the cycles into internal and external cycles. External cycles are those which appear in the closed process and involve edges which were absorbing edges in the original process. The set of all cycles, sorted into internal and external, for the example process fig~\ref{fig:example}, is shown in appendix~\ref{App:CycleList}. The external cycles correspond to the pathways from the starting state to an absorbing state. Therefore, the expected number of times traversing an external cycle before absorption will be at most one and corresponds to the probability of following a given path to absorption. Further, the sum of eqn.~\ref{eqn:cyclecount} over all external cycles will be one. 
\vspace{-1mm}

\subsection{Copolymer Methods}

\subsubsection{Philosophy of coarse-graining complex underlying copolymerisation reactions networks}
\label{method}
Armed with the techniques for solving absorbing Markov chains, here we set out the method for the analysis of copolymerisation processes.
Gaspard and Andrieux\cite{gaspard2014kinetics} presented a method to analyse Markov polymerisation processes in which each monomer is added in a single step (i.e. if the internal reaction network shown in figure~\ref{fig:modeltypes}d were trivial), assuming long polymers. We shall present a method for mapping more complex models for the individual polymerisation step onto coarse-grained descriptions that can be analysed using this framework, and then subsequently show how to back out the behaviour of the full model from the results.

 Consider a growing copolymer with $M$ monomer types, which are assumed to be present in the environment at fixed concentrations. At a coarse grained level, we can define a a state space of finite length sequences $\{x_1x_2\cdots x_l\;|\; x_i\in \{1,2,\cdots M\}, l\geq 0\}$, where $l$ is the length of the sequence. Let us refer to the coarse-grained states in this state space as \textit{completed states}.  On this coarse-grained level, a sequence of length $l$ may increase in length by one unit by polymerising one of $M$ units at the growing tip ($x_1x_2\cdots x_l \, \rightarrow \, x_1x_2\cdots x_lx_{l+1}$), or it may decrease in length by one unit ($x_1x_2\cdots x_l \, \rightarrow \, x_1x_2\cdots x_{l-1}$). Such a coarse grained model is depicted in figure~\ref{fig:modeltypes}(a,b,c) for free polymerisation, templated self-assembly and templated polymerisation with simultaneous separation.  

In general, copolymerisation processes may be best described by models in which the underlying copolymerisation reaction networks are complex, featuring multiple sub-steps in arbitrarily complex networks connecting the completed states, as suggested in figure~\ref{fig:modeltypes}d. Hence, overall, we could consider a copolymerisation process as having a tree-like structure with networks of reactions connecting completed states, as in figure~\ref{fig:petaldiagram}. Such a class of models is wide-reaching, with many examples from the literature included in this class\citeSolvable.

We will define a Markov process at the level of the coarse-grained completed states that, by construction, preserves probabilities of transitions between the completed states of the fine-grained process, and therefore preserves the statistics of the sequences produced. The coarse-grained Markov process does not preserve the distribution of transition times between completed states implied by the fine-grained model, which will in general be non-Markovian. Moreover, it does not provide fine-grained information on trajectories between the coarse-grained completed states. However, temporal details and information about the fine-grained dynamics can be added back in at a later stage, once statistics have been analysed at the coarse-grained level.

\subsubsection{Identifying propensities in the coarse-grained model}
\begin{figure*}
    \centering
    \includegraphics[scale=0.55]{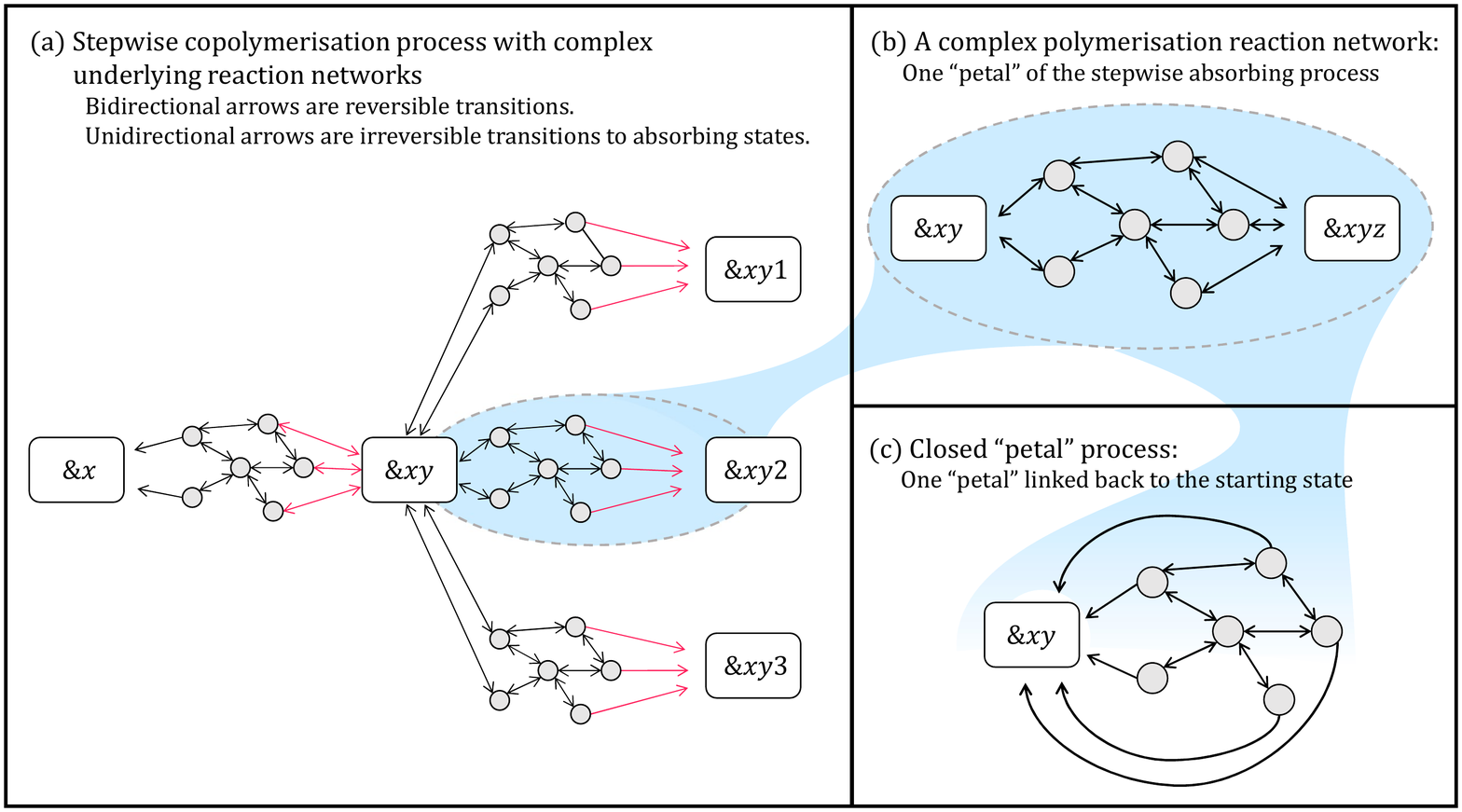}
    \caption{a) The step-wise process for an arbitrary model with 3 monomer types. This step-wise process is for a copolymer $\&xy$. The edges coloured red are the {\it completion} edges. The flower like structure of the step-wise process can be seen with 4 petals each connected at the starting state, $\&xy$. b) One of the petals of the step-wise process, which we use to define $\Lambda^\pm(z,y)$. $\Lambda^+(z,y)$ is defined as the sum of spanning trees rooted at the rightmost state, $\&xyz$ and $\Lambda^-(z,y)$ the sum of trees rooted at the leftmost state, $\&xy$. c) One of the petals (connecting $\&xy$ to $\&xyz$, as in b)) which has been linked back to the starting state. This graph is used to define $Q(z,y)$ as the sum of spanning trees rooted at the leftmost state, $\&xy$.}
    \label{fig:petaldiagram}
\end{figure*}

We find the transition rates  of the coarse-grained model (hereafter labelled propensities to avoid confusion with the underlying rates of the fine-grained process) by considering first passage problems between completed states. From a given completed state, there are $M+1$ completed states that may be reached, corresponding to the $M$ possible additions of a monomer and the removal of the monomer currently at the tip of the copolymer. For a first passage problem, we can convert each of these reachable completed states into an absorbing state by removing the transitions out of said states, as in figure~\ref{fig:petaldiagram}(a), in the same vein as Cady and Qian\cite{Cady2009Open}. Let us refer to this absorbing Markov process as the \textit{step-wise process} and define step to mean the addition/removal of a monomer.

We shall work with the assumption that the transition rates depend on the two monomers at the growing tip of the copolymer, following\citeTip. 
There will therefore be $M^2$ flavours of this process corresponding to the combinations of the two terminal monomers of the copolymer, the central state $\&xy$ (here $\&$ represents an arbitrary sequence). We wish to find the absorbing probabilities, $\mathbb{P}[\&xy\to\&xyz],\;z\in\{1,\cdots M\},\;\mathbb{P}[\&xy\to\&x]$ given an initial condition of the central state, $\&xy$. As outlined in the previous section, eqn.~\ref{eq:absprob}, we can find these probabilities by constructing the closed process and finding sums over spanning trees rooted at different states. The step-wise process has $M+1$ petal-like graphs each connected to the central state, but disconnected from each other. Due to this structure, any sums over spanning trees of the full process will factorise into a product of sums over spanning trees of the petals. Thus, we find that the absorbing probabilities take the following form:
\begin{eqnarray}
\nonumber
    \mathbb{P}[\&xy\to\&xyz]&=&\frac{1}{\mathcal{N}}\Lambda^+(z,y)\left[\prod_{z'\neq z}Q(z',y)\right]Q(y,x),\\
    \mathbb{P}[\&xy\to\&x]&=&\frac{1}{\mathcal{N}}\Lambda^-(y,x)\prod_{z}Q(z,y).
    \label{eq:lambda_and_Q}
\end{eqnarray}
Here $z\in\{1,\cdots M\}$, $\mathcal{N}$ is the normalisation factor from eqn.~\ref{eqn:MCTTss}; $\Lambda^+(z,y)$ is the sum over spanning trees of the petal connecting states monomers $\&xy$ and $\&xyz$, rooted at the forward completed state, $\&xyz$; $\Lambda^-(y,x)$ is the sum over spanning trees of the petal connecting states monomers $\&x$ and $\&xy$, rooted at the backwards completed state, $\&x$, figure~\ref{fig:petaldiagram}(b); and $Q(y,x)$ is the sum over spanning trees of the petal connecting states $\&x$ and $\&xy$, linked  back to the central state and rooted at the central state, i.e. with edges redirected to the starting state as in the closed process, as in figure~\ref{fig:petaldiagram}(c). Since $Q$ is a sum over spanning trees rooted at the node to which edges have been redirected, the sum takes the same form for both the forwards and backwards petals, only depending on which two completed states it is connecting.

From these probabilities, we see that choosing propensities $\omega_{\pm yx}$ for the transitions $\&x\xrightarrow[]{\omega_{+yx}}\&xy$ and $\&xy\xrightarrow[]{\omega_{-yx}}\&x$ such that
\begin{eqnarray}
    \omega_{\pm yx}=\frac{\Lambda^\pm(y,x)}{Q(y,x)}
    \label{eq:effectiverates}
\end{eqnarray}
not only preserves the ratios of probabilities of transitions to completed states, but also ensures that $\omega_{\pm yx}$ only depends on monomers $x$ and $y$.  We note here that this coarse graining process is different from lumping\cite{kemeny1983finite,esposito2012stochastic}, in which the state space is reduced while attempting to retain trajectory dynamics. In our approach, the coarse-grained process does not reproduce the dynamics of the fine-grained process, only the statistics of the completed states that are visited. However, dynamic quantities may be extracted exactly from the step-wise process, as we show in Sec.~\ref{Sec:ExtractingProp}.

\subsubsection{Solving the coarse-grained model}
We now use the methods developed by Gaspard and Andrieux\cite{gaspard2014kinetics} to solve the coarse-grained Markov model over the completed states, with propensities, $\omega_{\pm yx}$. Gaspard and Andrieux's approach considers a frame of reference that is comoving with the tip of the growing polymer, and assumes that the state of the tip and nearby monomers reaches a stationary distribution, to derive quantities at this steady state, such as the set of tip incorporation velocities, $v_x$ (the rates of adding monomers to a copolymer $\&x$),  the tip probabilities, $\mu(x)$ (the probability at a given time that the growing polymer is in state $\&x$), and the pair tip probabilities, $\mu(x,y)$ (the probability of being in state $\&xy$). However, we note that the time-dependent information is not physical at this stage due to the coarse-graining process. The above quantities are found from solving the following equations\cite{gaspard2014kinetics}:
\begin{eqnarray}
\label{eq:v}
    v_x&=&\sum_{y=1}^M\frac{\omega_{+yx}v_y}{\omega_{-yx}+v_y},\\\label{eq:mu}
    \mu(x)&=&\sum_{y=1}^M\frac{\omega_{+xy}}{\omega_{-xy}+v_x}\mu(y),\\
    \mu(x,y)&=&\frac{\omega_{+yx}}{\omega_{-yx}+v_y}\mu(x).
\end{eqnarray}
Using $\mu$ and $v$, we can calculate the statistics of the copolymer sequence far behind the growing tip \cite{gaspard2014kinetics}. We note that the distribution of monomers at the tip, $\mu(x)$, is different from the distribution of monomers at sites behind the tip; we assume that this distribution reaches some limit far behind the growing tip, in the bulk of the copolymer. This limiting distribution describes the probability that a monomer in the bulk of the copolymer takes a value $x$. Using $\varepsilon(x)$ to denote the frequency of monomer $x$ in the bulk of the copolymer,\cite{gaspard2014kinetics} 
\begin{eqnarray}
        \varepsilon(x)=\frac{\mu(x)v_x}{\sum\limits_y\mu(y)v_y}.
        \label{eq:bulk_prob}
\end{eqnarray} 
We may similarly define $\varepsilon(y|x)$ as the probability that in the bulk of the copolymer, a monomer $y$ is observed given a monomer $x$ behind it. $\varepsilon(x)$ and $\varepsilon(y|x)$ fully characterise the statistics of the bulk copolymer since under our assumptions - transitions only depend on the two monomers at the tip - the completed copolymer sequence is itself a Markov chain\cite{poulton2019nonequilibrium}.

 \subsubsection{Extracting properties of the fine-grained model from the solution of the coarse-grained model}
 \label{Sec:ExtractingProp}
The easiest quantities to extract are the frequencies of monomers in the bulk of the copolymer. These quantities are identical in the coarse-grained and fine-grained models, since the coarse-graining preserves the statistical distribution of sequences produced. Therefore $\varepsilon(x)$ as defined in eqn.~\ref{eq:bulk_prob} and $\varepsilon(y|x)$ apply directly to the fine-grained process.

The tip probabilities, $\mu$, above give the fraction of time spent in each tip state in the coarse-grained model. However, the coarse-grained model will not reproduce the time series of the fine-grained model, only the sequences of completed states visited. We therefore quotient out the lifetime of tip state $(x,y)$, $\tau(x,y)$, to obtain the frequency with which the tip states are visited in the coarse-grained model, 
\begin{eqnarray}
    \xi(x,y)&=&\frac{1}{\sum\limits_{x',y'=1}^M\frac{\mu(x',y')}{\tau(x',y')}}\frac{\mu(x,y)}{\tau(x,y)},\\
    \tau(x,y)&=&\frac{1}{\omega_{-yx}+\sum\limits_{z=1}^M\omega_{+zy}}.
    \label{eq:xi}
\end{eqnarray}
This frequency defines a new tip distribution, $\xi$. $\xi(x,y)$ is the frequency that a given pair of monomers $x,y$ is observed at the tip of the growing copolymer in the sequence of transitions. This distribution, $\xi(x,y)$, applies to both the coarse-grained model and the sequence of completed states visited in the full fine-grained model. It can therefore be used to find averages of key dynamic properties. 
 
For example, we can calculate the probability, $P$, that a growing copolymer increases in length at each step of the step-wise process. $P$ is calculated by averaging the probability of adding a monomer over the possible states $\&xy$:
\begin{eqnarray}
    P=\sum_{x,y=1}^M\xi(x,y)\frac{\sum\limits_{z=1}^M\omega_{+zy}}{\omega_{-yx}+\sum\limits_{z=1}^M\omega_{+zy}}.
    \label{eq:P}
\end{eqnarray}
Upon averaging out the sequence information we may treat the growth of a polymer as a random walk with probability $P$ of stepping forwards and $1-P$ of stepping back. We can find the expected number of monomer inclusion/removal steps per net forward step as $\nicefrac{1}{(2P-1)}$ (for proof see appendix~\ref{App:2P-1}). A number of quantities scale with the total number of steps rather than the net number of steps, making the number of steps per net forward step a necessary quantity. For example in order to find the expected time taken per net forward step, one can find the expected time to absorption for the step-wise process, figure~\ref{fig:petaldiagram}(a), $T(x,y)$, for a copolymer in state $\&xy$ by calculating $1/{J_{\rm Tot}(\&xy)}$ for the step-wise process using eqn.~\ref{eq:AbsorbingFlux}. The expected time per net forward step is then
\begin{eqnarray}
    \tau_{\text{step}} = \frac{1}{2P-1}\sum_{x,y=1}^M\xi(x,y)T(x,y).
    \label{eq:ExpectedTime}
\end{eqnarray} 
$1/{\tau_{\text{step}}}$ is therefore the physical average growth rate of the copolymer in the fine-grained model.

We may also calculate the chemical work done by the system in producing the copolymer. In a purely chemical system, with no time-varying externally applied protocols, the entropy increase of the universe is given by the decrease in the generalised free energy of the chemical system, including any coupled reservoirs of fuel molecules.\cite{ouldridge2018importance}. Since the total free energy must decrease, any increase in one contribution must be paid for by a decrease of at least the same magnitude in another contribution. It is common to describe the latter subsystem as doing work on the former. 

For the polymerisation systems analysed here, the generalised free energy can be split into a term corresponding to the chemical free energy of the system, averaged over the uncertain state of the system, and a term related to the entropy arising due to the uncertainty of the state occupied.\cite{ouldridge2019power}
\begin{eqnarray}
    \mathcal{G} = \sum_a p(a) {G}_{\rm chem} (a) + \sum_a p(a) \ln p(a),
    \label{eq:generalized_fe}
\end{eqnarray}
where we use natural units such that $k_B T=1$. Here, $a$ is a chemical state of the system as a whole,  ${G}_{\rm chem} (a)$ is the chemical free energy of state $a$, and $p(a)$ is the probability that the system occupies the state $a$. ${G}_{\rm chem} (a) = -\ln Z_a$, where $Z_a$ is the partition function of the system (explicitly including any large chemical buffers) restricted to the chemical state $a$, and represents the contribution of concentrations and bond strength to the favourability of a molecular state. The principle of detailed balance\cite{ouldridge2018importance} states that the chemical free energy change associated with a transition from $a$ to $b$ is given by
\begin{eqnarray}
         {G}_{\rm chem} (b) - {G}_{\rm chem} (a)=-\ln\left(\frac{K(a,b)}{K(b,a)}\right).
        \label{eq:edge_dG}
\end{eqnarray}

The second term in eqn.~\ref{eq:generalized_fe} is information theoretic in character; it is equal to the negative of the Shannon entropy associated with the distribution over chemical states. For the systems studied here, in which we consider infinitely long copolymers that have reached steady state growth, the only relevant contribution to this term is the increase in Shannon entropy of the copolymer sequence produced as the polymer gets longer. Since the copolymer sequence is itself a discrete time Markov chain\cite{poulton2019nonequilibrium} the additional entropy per net forward step (the entropy rate) can be readily calculated \cite{cover2006elements}:
\begin{eqnarray}
        H=-\sum_{x,y=1}^{M}\varepsilon(x)\varepsilon(y|x)\ln\varepsilon(y|x),
        \label{eq:seqentropy}
\end{eqnarray}
with $x,y$ representing the monomer types. Since the purpose of a copolymerisation system is often to produce a low entropy (or ``accurate'') sequence, it is reasonable to think of the chemical free-energy decrease per net forward step as the chemical work done to reduce the information entropy of eqn.~\ref{eq:seqentropy} below that of a uniform, random polymer.   Extending the definition provided by Poulton et al.\cite{poulton2019nonequilibrium}, we may define the efficiency of copolymerisation as:
\begin{eqnarray}
\eta=\frac{\ln M-H}{\ln M +\mathcal{W}_{\rm chem}}\leq1,
\label{eq:efficiency}
\end{eqnarray}
$\ln M$ is the entropy per monomer (or entropy rate) of a uniform, random copolymer with $M$ monomer types, and $\mathcal{W}_{\rm chem}$ is the average decrease in chemical free energy per net forward step. This efficiency is then ratio between the entropy drop due to the accuracy of the copolymer compared to a random one ($\ln M-H$) and the chemical work used to drive the system ($\mathcal{W}_{\rm chem}$) above that required to make a random copolymer in equilibrium (-$\ln M$)\cite{EspositoExtracting}.

The expected work done during a transition adding or removing a monomer given starting in completed state $\&xy$ can be calculated by summing the contribution from eqn.~\ref{eq:edge_dG} multiplied by the expected net current along the edge $a \leftrightharpoons b$ prior to absorption over all edges in the step-wise process:
\begin{eqnarray}
        w_{\rm chem}(x,y) &=& \\ -\Delta G_{\rm chem}(x,y) &=& \sum_{b>a}\ln\left(\frac{K(a,b)}{K(b,a)}\right)N_{a\leftrightharpoons b}(\&xy),
        \label{eq:edgework}
\end{eqnarray}
where $N_{a\leftrightharpoons b}(\&xy)$ is the expected net number of times traversing edge $a\leftrightharpoons b$ before absorption giving starting in the central state of the step-wise process, $\&xy$, as in eqn.~\ref{eq:edgecount}. This sum will also require  contributions from edges which lead to absorbing states. For such edges, the rate for the reverse transition in the logarithm of eqn.~\ref{eq:edgework} is the rate from the full process.

Equivalently, however, as outlined in appendix~\ref{App:Edge-Cycle}, we may find this chemical work by considering the non-recurrent cycles of the process\cite{Wachtel2018Thermodynamically}. For a given internal cycle, $C$, we may define the affinity\cite{ouldridge2018importance}, 
\begin{eqnarray}
    A(C)=\ln\frac{\prod\limits_{e\in C}K(e)}{\prod\limits_{e\in C'}K(e)},
    \label{eq:affinity}
\end{eqnarray}
where the sum is over the edges, $e$, composing the cycle and $C'$ is the cycle with edges in revered direction. For external cycles, we may define the affinity in the same way, inferring the rate for the reversed edge of the transition to absorbing states from the full process. The expected work done before absorption of the cycle, $C$, given starting in the state $\&xy$ is 
\begin{eqnarray}
      w_{\rm chem}(x,y) &=& \\
      -\Delta G_{\rm chem}(x,y) &=& \sum_C A(C)\frac{J_{\text{Cyc}}(\sigma,C)-J_{\text{Cyc}}(\sigma,C')}{J_{\text{Tot}(\&xy)}}.
    \label{eq:cycleenergy}
\end{eqnarray}
Averaging $ w_{\rm chem}(x,y)$ with $\xi$ and multiplying by the expected number of steps per net forward step gives the expected chemical work done per net forward step,
\begin{eqnarray}
    \mathcal{W}_{\rm chem} = \frac{1}{2P-1}\sum_{x,y=1}^M\xi(x,y) w_{\rm chem}(x,y).
   \label{eq:FreeEnergyLoss}
\end{eqnarray}

Further, the forms of eqns.~\ref{eq:ExpectedTime} and \ref{eq:FreeEnergyLoss} may be applied to an arbitrary quantity for which one can find the expected value in the step-wise process starting in state $\&xy$. Let this arbitrary quantity be $A(x,y)$. One can then average this using the distribution, $\xi$, to obtain the expected value of the quantity per step. Then, if appropriate, multiplying by $\nicefrac{1}{(2P-1)}$, gives the expected value of the quantity per net forward step. In practice, as shall be seen in section~\ref{sec:KP}, since the quantities we wish to calculate may be written in terms of sums over spanning trees, the quantities for the step-wise process may be written as a sum over the terms per petal, with the quantity for a given petal factorising into some quantity which depends on the petal multiplied by $Q$'s for the other petals.

\subsubsection{Stalled growth}
\label{sec:Stall}
Explicit simulation of copolymer growth is particularly challenging in regimes where $P \gtrsim 0.5$, since many backward and forwards steps are taken per net forwards step. At $P= 0.5$, then the process will not reliably produce copolymers; for $P<0.5$ polymers will tend to shrink. In general, for $P = 0.5$, we can say the model has stalled. Our approach is particularly beneficial in this case; indeed, it is  possible to check whether a model is at the stall point by considering an $M \times M$ dimensional matrix of the ratios of forward to backwards propensities\cite{gaspard2014kinetics}, $Z_{yx}=\left(\frac{\omega_{+yx}}{\omega_{-yx}}\right)=\frac{\Lambda^+(y,x)}{\Lambda^-(y,x)}$. The model is at the stall point if and only if:
\begin{eqnarray}
\det\left(Z-\mathbbm{1}_M\right)=0,
\label{eq:stallcondition}
\end{eqnarray}
where $\mathbbm{1}_M$ is the $M\times M$ identity matrix, and shrinking if negative. Since $Z$ gives the ratios of adding a monomer to removing one, this condition essentially says that models will stall if the total rate of adding a monomer is equal to the total rate of removing one. 

In a typical model, there exists at least one parameter that controls the driving. Often this parameter is related to the backbone strength of the polymer produced: {\it e.g.} the free energy drop associated with the formation of a generic backbone bond, $\Delta G_{\pol}$. This parameter will be present in the rates of each external cycle so that by tuning it, the model can be moved all the way from stalling to irreversible driving, whereby monomers cannot be removed once polymerised. If such a parameter exists, we may rephrase the stall condition, eqn.~\ref{eq:stallcondition}, in terms of this parameter. For example, for the case of the parameter being $\Delta G_{\pol}$, we may find some threshold value $\Gamma$ such that the model will stall for $\Delta G_{\pol}=\Gamma$.
 
 \subsubsection{Limiting behaviour}
We shall note two limits for which we may give analytic expressions for the frequency of monomer types in the copolymer bulk  for all models. First, consider the case that the system is at the stall point (eqn.~\ref{eq:stallcondition}). In general, entropy production can still occur within cycles in the step-wise process; therefore, these frequencies cannot be determined from equilibrium arguments and are non-trivial. Nonetheless, at the stall point, we may express the monomer frequencies in the bulk relatively simply. The frequency of monomer $x$, $\varepsilon_{\stall}(x)$, is proportional (up to normalisation) to the cofactor of the diagonal element (corresponding to monomer $x$) of the matrix $\left(\mathbbm{1}_M-Z\right)$, as proven in appendix~{\ref{App:Estall}}. For example, for $M=2$,
\begin{eqnarray}
\nonumber
       \varepsilon_{\stall}(1)&\propto&1-\frac{\omega_{+22}}{\omega_{-22}},\\\varepsilon_{\stall}(2)&\propto&1-\frac{\omega_{+11}}{\omega_{-11}},
       \label{eq:StallError}
\end{eqnarray}
 and for $M=3$, we have
\begin{eqnarray}
\nonumber
       \varepsilon_{\stall}(1)&\propto&\left(1-\frac{\omega_{+22}}{\omega_{-22}}\right)\left(1-\frac{\omega_{+33}}{\omega_{-33}}\right)-\frac{\omega_{+23}}{\omega_{-23}}\frac{\omega_{+32}}{\omega_{-32}},\\
\nonumber
       \varepsilon_{\stall}(2)&\propto&\left(1-\frac{\omega_{+11}}{\omega_{-11}}\right)\left(1-\frac{\omega_{+33}}{\omega_{-33}}\right)-\frac{\omega_{+13}}{\omega_{-13}}\frac{\omega_{+31}}{\omega_{-31}},\\
   \varepsilon_{\stall}(3)&\propto&\left(1-\frac{\omega_{+11}}{\omega_{-11}}\right)\left(1-\frac{\omega_{+22}}{\omega_{-22}}\right)-\frac{\omega_{+12}}{\omega_{-12}}\frac{\omega_{+21}}{\omega_{-21}}.
\end{eqnarray}

On the other end of the spectrum, we can also solve for monomer bulk frequencies in the irreversible limit, where $\omega_{-yx}=0$ for all $x,y$. Intuitively, we could consider the Markov process on the state space $\{1,\cdots,M\}$ representing copolymers with a given monomer at its tip, and transitions between those states with rates, $K_{\irr}(x\to y)=\omega_{+yx}$. The steady state of this process will give the time dependent frequencies of having a given monomer at the tip of the copolymer. Therefore, dividing by the time spent in each state will give the bulk frequencies. A nice way to write out these frequencies in the style of the methods described thus far is as a sum over the spanning trees on the complete graph on $M$ vertices with rate functions $K_{\irr}(x,y)=\omega_{+yx}$. Explicitly, we may write these frequencies (up to normalisation) as:
\begin{eqnarray}
        \varepsilon_{\irr}(x)\propto \left(\sum\limits_{T\in\mathcal{T}(x)}\prod\limits_{e\in T}K_{\irr}(e)\right)\sum_{y=1}^M\omega_{+yx},
        \label{eq:IrreversibleError}
\end{eqnarray}
where $\mathcal{T}(x)$ is the set of spanning trees of the complete graph on $M$ vertices. This expression is derived  formally in appendix~\ref{App:Eirrev}. For example, with $M=2$,
\vspace*{-0.5mm}
\begin{eqnarray}
        \nonumber
        \varepsilon_{\irr}(1)&=&\frac{\omega_{+12}(\omega_{+11}+\omega_{+21})}{\omega_{+12}(\omega_{+11}+\omega_{+21})+\omega_{+21}(\omega_{+12}+\omega_{+22})},\\ \nonumber
        \varepsilon_{\irr}(2)&=&\frac{\omega_{+21}(\omega_{+12}+\omega_{+22})}{\omega_{+12}(\omega_{+11}+\omega_{+21})+\omega_{+21}(\omega_{+12}+\omega_{+22})}.\\
\vspace{-4mm}
\end{eqnarray}

\vspace{-4mm}
\subsubsection{Simplification for factorisable propensities}
\vspace{-2mm}
The presented method applies to arbitrary complex copolymerisation models obeying the structure of figure~\ref{fig:petaldiagram}. However, if we make some further common assumptions, much of the analysis simplifies. 
For example, consider the case in which the ratios of propensities may be factored:
\begin{eqnarray}
\frac{\omega_{+yx}}{\omega_{-yx}}=\frac{\Lambda^+(y,x)}{\Lambda^-(y,x)}=Y(y)X(x),
\label{eq:FactorisationRates}
\end{eqnarray}
where $Y$ is a function of monomer $y$ only and $X$ is a function of monomer $x$ only. Intuitively, such a condition holds in the cases where there is no direct, type-dependent interactions between monomers in the growing polymer, such as when monomers only interact with a template\citeTemp. Under such an assumption, multiple calculations simplify, see appendix~\ref{App:factorisable}. For example, the stall condition becomes simply that the model will stall at
\begin{eqnarray}
\sum_xX(x)Y(x)=1,
\label{eq:FactorisedStallCondition}
\end{eqnarray}
Bulk frequencies at stall are just:
\begin{eqnarray}
\varepsilon_{\stall}(x)=X(x)Y(x).
\label{eq:FactorisedStallError}
\end{eqnarray}

\section{Example Applications}
\label{examples}
We shall now consider some exemplar classes of models to: provide examples of how to utilise the methods; validate their accuracy; and to show the types of quantities and information that may be extracted. 

A useful initial classification of models is into those which we shall call balanced. We shall refer to a model as being balanced if its petals (see figure~\ref{fig:petaldiagram}(b)) are detailed balanced. Such models are useful baseline checks as their cycles all have zero affinity, meaning no chemical work is done internally and hence the only contributions to chemical work are from external cycles. Further, these models exhibit a proper equilibrium at the stall point, and as such allow for equilibrium arguments to validate the method at this point. It is worth noting that although related to the notion of detailed balanced, the full model with its infinite state space is not detailed balanced.

\subsection{Stalling behaviour in a polymerisation model with no neighbour-neighbour interactions}
We shall start with the simplest case, where the propensities in the coarse-grained model only depend on the monomer type being added/removed: $\omega_{\pm yx}=\omega_{\pm y}$, such as in a simple model for templated self assembly, figure~\ref{fig:modeltypes}(b). Assume there exists a backbone free energy, $\Delta G_{\pol}$ controlling the driving as in section~\ref{sec:Stall}. Any spanning tree in $\Lambda^\pm$ must involve at least one incidence of $\Delta G_{\pol}$, since it appears in every external cycle. Therefore, we can split the ratio of propensities as follows:
\begin{eqnarray}
\frac{\omega_{+y}}{\omega_{-y}}=e^{\Delta G_y}e^{\Delta G_{\text{pol}}},
\end{eqnarray}
where $\Delta G_y$ encompasses the rest of the details about the models. We note in general, $\Delta G_y$ may be a function of $\Delta G_{\pol}$, however in many cases, it is not. These cases include when there is only one completion reaction (highlighted in red in figure~\ref{fig:petaldiagram}(a)) that contains the dependence on $\Delta G_{\pol}$ or if the model is balanced. We may then interpret $-\Delta G_y$ as an effective binding free energy of monomer $y$. If we think of $\Delta G_{\pol}$ as the free energy drive of the model away from stall, we look for a threshold value $\Delta G_{\pol}=\Gamma$ above which the model will not stall. Using eqn.~\ref{eq:FactorisedStallCondition}, we see that
\begin{eqnarray}
\Gamma=-\ln \left(\sum_y e^{\Delta G_y}\right)=-\ln\mathcal{Z},
\end{eqnarray}
where $\mathcal{Z}$ is the partition function for a system with one state for each monomer type, each state labelled by $y$ and with free energy $-\Delta G_y$.  Furthermore, using eqn.~\ref{eq:FactorisedStallError}, the bulk frequencies at the stall point may be written:
\begin{eqnarray}
\varepsilon_{\stall}(y)=\frac{e^{\Delta G_y}}{\sum_xe^{\Delta G_x}}=\frac{1}{\mathcal{Z}}e^{\Delta G_y},
\end{eqnarray}
which is the probability of selecting a state $y$, with free energy, $-\Delta G_y$ as predicted by equilibrium statistical mechanics. In these results, $-\Delta G_y$ looks like the equilibrium contribution to free energy, and the results follow fairly directly in equilibrium. However, these results hold even if the process involves fuel-consuming cycles: entropy may still be being produced at stall. In such cases, the effect of breaking equilibrium will be to change the effective free energies of selecting a given monomer type. 

\subsection{Balanced models of templated polymerisation with autonomous separation}
\label{sec:DB_pol_sep}
Next we shall consider a class of models where the ratio of propensities may be written:
\begin{eqnarray}
\frac{\omega_{+yx}}{\omega_{-yx}}=e^{\Delta G_y}e^{-\Delta G_x}e^{\Delta G_{\pol}}.
\label{eq:DBASRatio}
\end{eqnarray}
As before, $\Delta G_{\pol}$, coming from the polymerisation reactions represents the driving of this process. Such a class of models includes, most notably, balanced models of templated polymerisation with autonomous separation,\cite{poulton2019nonequilibrium}. In these cases the breaking of the previous copy-template bond every time a new bond is formed enforces the structure in eqn.~\ref{eq:DBASRatio}. We shall assume, as in Ref.~\onlinecite{poulton2019nonequilibrium}, that  $\Delta G_y$ is independent of $\Delta G_{\pol}$. 

Using eqn.~\ref{eq:FactorisedStallCondition} and eqn.~\ref{eq:FactorisedStallError}, we find the stall point to be $\Delta G_{\pol}=\Gamma=-\ln M$ and bulk frequencies at stall, $\varepsilon_{\stall}(y)=\frac{1}{M}$, where $M$ the number of monomer types. Physically, we can understand these results by considering balanced models of templated polymerisation with autonomous separation. For such models, by definition there is no entropy production in internal cycles and therefore, the stall point must be thermodynamic equilibrium. In such models, the only driving comes from the polymerisation, $\Delta G_{\pol}$, and the entropic effect  having $M$ monomers to choose. These two effect balance at equilibrium.\cite{EspositoExtracting}

Next let us consider the limit that the completion reactions highlighted in red in figure~\ref{fig:petaldiagram}\,(a) are much slower than the other reactions. Explicitly, let $k$ be some rate constant at the same order of magnitude of the rates of the process that are not the rates for the completion transitions indicated in red in figure \ref{fig:petaldiagram}\,(a). Write the completion rates as $k_{\abs}R_\abs^+(y,x)$, where $k_{\abs}\ll k$ is a rate constant controlling the overall speed of the completion reactions and $R_\abs^+(y,x)$ provides any sequence dependence. Similarly, the reverse transitions along the completion edges have the rate $k_{\abs}R_\abs^-(y,x)$. Further, let there be $n_{\abs}$ such completion reactions in a given petal of the step-wise process (we shall assume this number is the same for all pairs of monomers, $x,y$) 

Assume for simplicity that all completion reactions, $R_\abs^\pm(y,x)$, take the same form in a given petal. Then, we can write the sum over spanning trees, $Q(y,x)$ as 
\begin{eqnarray}
Q(y,x)=\frac{1}{n_{\abs}}\frac{\Lambda^-(y,x)}{k_{\abs}R_\abs^-(y,x)}+\mathcal{O}\left(\frac{k_{\abs}}{k}\right),
\end{eqnarray}
since $\Lambda^-(y,x)$ has first order terms in ${k_{\abs}}/{k}$. This fact can be seen from noting that the leading order terms in $Q(y,x)$ are the trees with no completion reactions and the leading order terms in $\Lambda^-(y,x)$ are those same leading order trees of $Q(y,x)$, except with one completion reaction added in. There are $n_{\abs}$ such completion reactions and each adds the same leading order term to $\Lambda^-(y,x)$. With $Q(y,x)$ taking this form, and remembering eqn.~\ref{eq:DBASRatio}, the propensities take the following form:
\begin{eqnarray}
\nonumber \omega_{+yx}&=&n_{\abs}k_{\abs}R^-_{\abs}(y,x)e^{\Delta G_y-\Delta G_x+\Delta G_{\pol}}+\mathcal{O}\left(\frac{k_{\abs}}{k}\right)^2,\\
\omega_{-yx}&=&n_{\abs}k_{\abs}R^-_{\abs}(y,x)+\mathcal{O}\left(\frac{k_{\abs}}{k}\right)^2.
\label{eq:DBProp}
\end{eqnarray}
The $n_{\abs}k_{\abs}$ term cancels in ratios of $\omega_{\pm yx}$ variables, and therefore does not affect the sequence statistics. Thus, in the slow completion limit, such models are only affected by the binding free energy differences ($\Delta G_y-\Delta G_x$), the driving ($\Delta G_{\pol}$), and the nature of the final completion step ($R^-_{\abs}$). Therefore, the fine details do not affect the statistics of the polymers.

\begin{figure}
    \centering
    \includegraphics[scale=0.5]{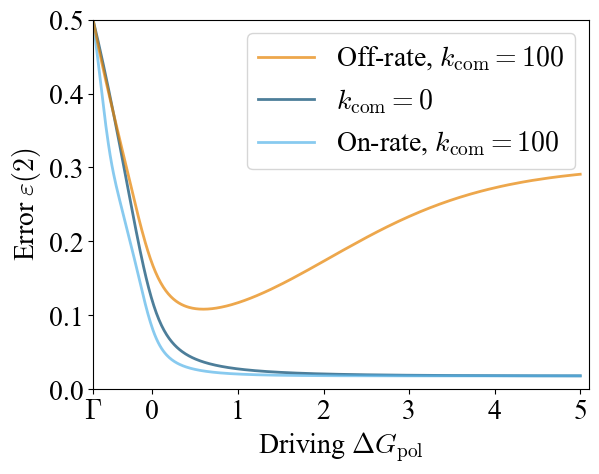}
    \caption{Plots of the frequency of the less stably-bound (incorrect) monomer with smallest binding free energy, labelled $2$ for on- and off-rate discrimination balanced models with $k_{\abs}=100$ and $k_{\abs}\rightarrow 0$. The binding free-energy difference for these models is $DG=\Delta G_1-\Delta G_2=4$. The models are topologically the Hopfield model as in figure~\ref{fig:HopfModel}a, with $\Delta G_{\act}=0$ and $M_{\ina}=M_{\act}=1$. However, for the on-rate discrimination, the free-energy terms are in the binding reactions instead of the unbinding ones. The specific models are given in appendix~\ref{App:DBModels}.}
    \label{fig:onoffdiscrimination}
\end{figure}

Assuming that all completion edges are associated with the same free energy change $-\Delta G_{\rm pol}$, so that  $R^-_{\abs}(y,x)=e^{-\Delta G_{\pol}}$, we may solve for the statistics explicitly. For the case of two monomer types, $M=2$, we find the bulk frequency to be (appendix~\ref{App:SlowPolyDB}): 
\begin{eqnarray}
    \nonumber\varepsilon(1)&=&\bigg(1-\frac{1}{2}(e^{-\Delta G_{\pol}}-1)(e^{-DG}-1)\\
    &+&\frac{1}{2}\sqrt{(e^{-\Delta G_{\pol}}-1)^2(e^{-DG}-1)^2+4e^{-DG}}\bigg)^{-1},
    \label{eq:DBErr}
\end{eqnarray}
where $DG=\Delta G_1-\Delta G_2$. This expression is plotted in figure~\ref{fig:onoffdiscrimination} for $DG=4$. From this expression, we can confirm explicitly by substituting in the stall driving, $\Delta G_{\pol}=-\ln2$, that the bulk frequency indeed becomes $\varepsilon(1)=\frac{1}{2}$. Further, taking the irreversible limit, $\Delta G_{\pol}\to\infty$, we find the bulk frequency becomes:
\begin{eqnarray}
\varepsilon(1)=\frac{e^{\Delta G_1}}{e^{\Delta G_1}+e^{\Delta G_2}},
\end{eqnarray}
the equilibrium statistical mechanics probability of choosing state $1$ with free energy $-\Delta G_{1}$, given state 2 has free energy $-\Delta G_2$. Since the completion reactions are slow and irreversible, in this limit, the process selecting the monomers is allowed to equilibriate. Therefore, copolymerisation is simply sampling from the equilibrium distribution of this process, and hence tends to the result predicted by equilibrium statistical mechanics. 
 
Eqn.\ref{eq:DBProp} shows that in the slow completion limit, the fine details of the reaction network leading to selection of a specific monomer become unimportant and the models collapse onto a single accuracy curve determined by $DG$, $\Delta G_{\pol}$ and $R^-_{\abs}$. Conversely, if we fix all parameters except $k_{\abs}$, we seem to see that the bulk frequencies will tend monotonically to their limits as ${k_{\abs}}/{k}\to0$, either from above or below.

We can use this fact to compare bulk frequencies for certain types of model. For example, we may compare on-rate discrimination,\cite{sartori2013kinetic} where incorrect monomers bind more slowly, to off-rate discrimination,\cite{sartori2013kinetic} where incorrect monomers unbind more quickly. An example model comparing on-rate and off-rate discrimination is plotted in figure~\ref{fig:onoffdiscrimination} for a model defined in appendix~\ref{App:DBModels}. Consider the bulk frequency of an incorrect monomer. On-rate discrimination benefits from fast polymerisation and therefore tends to its slow polymerisation limit from below, whereas off-rate discrimination benefits from allowing the process selecting monomers to equilibriate and hence tends to its slow copolymerisation limit from above. This fact sets up a hierarchy for a given set of parameters, and moderate or strong driving, for the bulk frequency of incorrect monomers, off-rate discrimination > slow copolymerisation > on-rate discrimination. This observation is consistent with the results of Sartori and Pigolotti~\cite{sartori2013kinetic} and Poulton et al.\cite{poulton2019nonequilibrium} for kinetic (on-rate) and energetic (off-rate) discrimination.

\subsection{Hopfield's Kinetic Proofreading in a model of templated copying with autonomous separation}
\label{sec:KP}
For our final example, we shall consider an explicit model of copolymerisation, with Hopfield's kinetic proofreading mechanism incorporated into a templated copolymerisation system with autonomously separating product in a thermodynamically valid way. From this setup, we can provide a fully worked example of an explicit model, as well as demonstrating the power of the method for analysing sequences of models with recursive structures as we look at a generalised version of Hopfield's proofreading incorporated into a model of templated polymerisation with autonomous separation. 

\begin{figure*}
    \centering
    \includegraphics[scale=0.5]{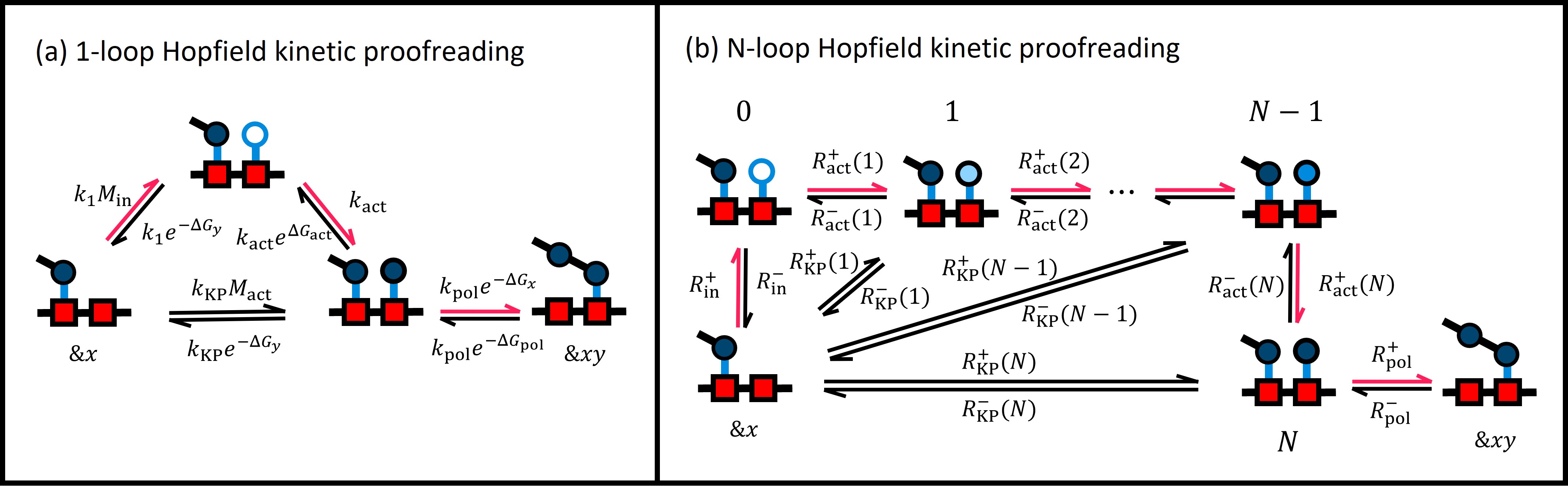}
    \caption{Reaction rates of the (a) $1$-loop and (b) $N$-loop Hopfield kinetic proofreading models implemented in a templated polymerisation model with autonomous separation system. Each of these subfigures represents a single petal of the step-wise process as in figure~\ref{fig:petaldiagram}, going from completed state $\&x\to\&xy$. In both cases, the template is represented by red squares. In (a) the inactive monomer is represented by a white circle and the activated monomers by a dark blue circle. In (b), different levels of activation are represented by increasingly dark shades of blue circles. Further, in (b) the numbers by the states represent the activation level of the monomer. In each case, the desired pathway is highlighted with red arrows.}
    \label{fig:HopfModel}
\end{figure*}

Explicitly, we first consider the one-loop model of kinetic proofreading shown in figure~\ref{fig:HopfModel}\,(a). There are two monomer types, the right ones $x=r$ and wrong ones $x=w$. Note that we have already transformed the model so that the sequence of the copy is defined relative to that of the template\cite{bennett1979dissipation}. These monomer types exist in inactive and active states with concentrations $M_{\ina}$ and $M_{\act}$, respectively, relative to some reference concentration, with each monomer type having the same concentration. As previously, we shall assume the environment is sufficiently large such that these concentrations remain constant.

The monomers may bind to the template either in an active or inactive state with binding free energies $-\Delta G_x$ for monomer type $x$. Inactive monomers may be activated on the template with a free-energy change of $\Delta G_{\act}$. Finally, active monomers may be polymerised into the copolymer chain, with free-energy change $-\Delta G_{\pol}$. Subsequently, the penultimate monomer of the copolymer unbinds from the template. Each of these reactions is assigned a forwards and reverse reaction rate consistent with the thermodynamic model; the full model is illustrated in figure~\ref{fig:HopfModel}\,(a). Conceptually, the proofreading motif functions by providing two opportunities to reject the unwanted monomer $w$: first, when the un-activated monomer binds, and second, after it has been activated. To be effective, a non-zero affinity is required to drive the system around the cycle of states in the correct order: unbound template site $\rightarrow$ unactivated monomer bound $\rightarrow$ activated monomer bound.\cite{hopfield1974kinetic,ouldridge2018importance} We emphasise that this model differs from Hopfield's original description in two important ways: firstly, we consider a full,  microscopically reversible polymerisation process, rather than a single incorporation step with irreversible polymerisation; and secondly, we embed the proofreading motif into a non-trivial polymerisation process involving autonomous detachment from the template.

Given the model as described in figure~\ref{fig:HopfModel}\,(a), we first identify the propensities $\omega_{xy}$ connecting completed states. Due to the petal-like structure, we can follow eqn.~\ref{eq:lambda_and_Q} and simply consider spanning trees of the petal sub-processes illustrated in figure~\ref{fig:HopfModel}; $\Lambda^-(y,x)$ rooted at $\&x$, $\Lambda^+(y,x)$ rooted at $\&xy$, and $Q(y,x)$, for a petal connecting $\&x$ and $\&xy$. Explicitly writing out the sums of spanning trees, we obtain:
\onecolumngrid
\begin{eqnarray}
\Lambda^+_1(y,x)&=&\left[k_1k_{\act}M_{\ina}+k_{KP}M_{\act}(k_1e^{-\Delta G_y}+k_{\act})\right]k_{\pol}e^{-\Delta G_x},\\
\Lambda^-_1(y,x)&=&\left[k_1k_{\act}e^{\Delta G_{\act}-\Delta G_y}+k_{KP}e^{-\Delta G_y}(k_1e^{-\Delta G_y}+k_{\act})\right]k_{\pol}e^{-\Gpol},\\
Q_1(y,x)&=& \bigg[k_1k_{\act}e^{\Delta G_{\act}-\Delta G_y}+(k_{KP}e^{-\Delta G_y}+k_{\pol}e^{-\Delta G_x})(k_1e^{-\Delta G_y}+k_{\act})\bigg].
\end{eqnarray}
\twocolumngrid
Here, we add a subscript $1$ to denote these as for the simple, "1-loop", Hopfield model, which we shall extend to allow more loops later. We note that the ratio, ${\Lambda^+(y,x)}/{\Lambda^-(y,x)}$ factorises as eqn.~\ref{eq:FactorisationRates} and so we can easily write down the stall condition as $\Gpol=\Gamma$ with
\begin{eqnarray}
\nonumber\Gamma&=&-\ln\bigg(\frac{k_1 k_{\text{KP}} M_{\act}e^{-\Delta G_r}+k_{\act}(k_{\KP}M_{\act}+k_1M_{\ina})}{k_1k_{\KP}e^{-\Delta G_r}+k_{\act}(k_1e^{\Delta G_{\act}}+k_{\KP})}\\
&+&\frac{k_1 k_{\text{KP}} M_{\act}e^{-\Delta G_w}+k_{\act}(k_{\KP}M_{\act}+k_1M_{\ina})}{k_1k_{\KP}e^{-\Delta G_w}+k_{\act}(k_1e^{\Delta G_{\act}}+k_{\KP})}\bigg).
\label{eq:hopfGam}
\end{eqnarray}
Note that setting $M_{\ina}=M_{\act}=1,\Delta G_{\act}=0$, in eqn.~\ref{eq:hopfGam}, $\Gamma$ collapses to $-\ln2$ as these conditions reduce the system to a balanced one with a stall point at equilibrium, as in Section~\ref{sec:DB_pol_sep}.

\begin{figure}
    \centering
    \begin{subfigure}{0.48\textwidth}
    \includegraphics[scale=0.5]{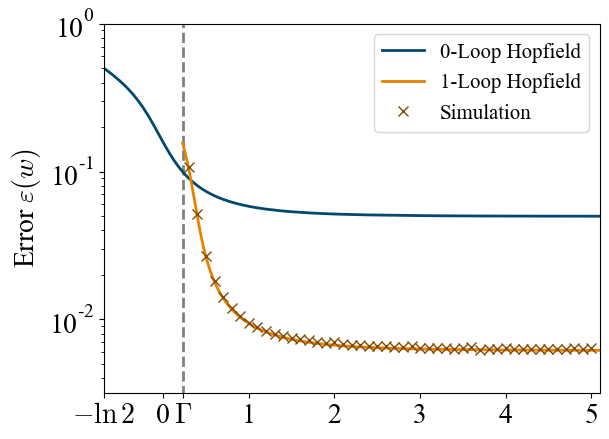}
    \label{fig:HopfError}
    \end{subfigure}
    
    \begin{subfigure}{0.48\textwidth}
    \includegraphics[scale=0.5]{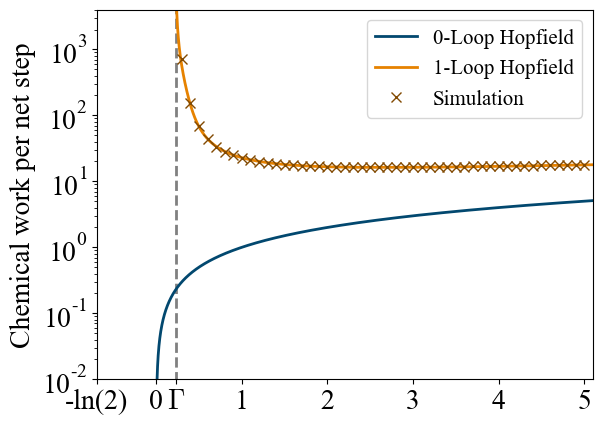}
    \label{fig:HopfEnergy}
    \end{subfigure}
    
    \begin{subfigure}{0.48\textwidth}
    \includegraphics[scale=0.5]{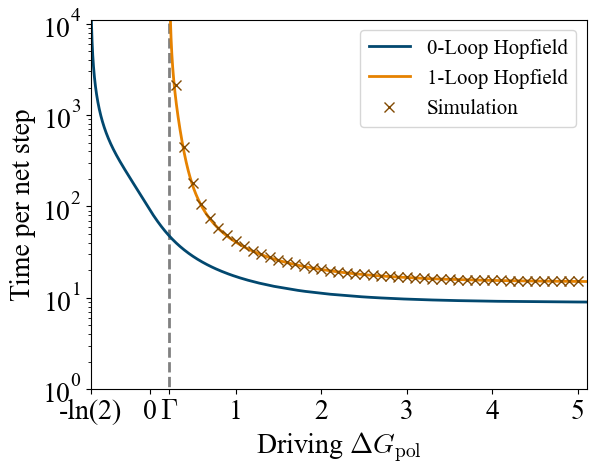}
    \label{fig:HopfTime}
    \end{subfigure}
    \caption{The analytical method applied to a 1-loop proofreading model (figure~\ref{fig:HopfModel}\,(a)), compared to Gillespie simulation of the same model and a simpler 0-loop model. For these data, the following parameters were used: $\Delta G_r=2,\;\Delta G_w=-2,\;\Delta G_{\act}=-1\; M_{\ina}=1,\;M_{\act}=0.01,\;k_1=k_{\act}=k_{\KP}=1$. The stall point, $\Gamma$, is marked on each of the plots. The Gillespie simulations used a template of length 2000 and were run till completion with the first monomer being chosen as either $r$ or $w$ with probability $0.5$. The statistics were averaged over 2000 copolymers per data point. The chemical work was calculated from the simulation as $(\text{Inactive monomers})*(\Delta G_{\act}-\ln (M_{\act}/M_{\ina}))+L*(\Delta G_{\pol}+\ln M_{\act})$ where ``Inactive monomers" is the number of inactive monomers taken out of the environment and $L$ is the length of the template.}
    \label{fig:HopfData}
\end{figure}

The frequency of right and wrong monomers, $\epsilon(x=r,w)$ may be calculated from eqn.~\ref{eq:bulk_prob} (the calculation is implemented in the supporting information). We plot copying error, as represented by $\epsilon(w)$, in figure~\ref{fig:HopfData}\,(a), and demonstrate that it agrees well with the results found from a Gillespie simulation\cite{gillespie1976general} of the same model. We also compare to a ``0-loop'' version of the model, in which the inactivated monomers and the inactivated monomer bound state are omitted. As can be seen, the proofreading motif generally improves accuracy when driven above its stall point $\Delta G_{\rm pol}= \Gamma$.  Indeed, we may write down expressions for the bulk frequency in the irreversible limit ($\Delta G_{\pol}\to\infty$) using eqn.~\ref{eq:IrreversibleError}. In this irreversible limit, we recover Hopfield's classic argument by taking some further limits consistent with his analysis. Namely, letting $M_{\act},k_{\act},k_{\pol}\to 0$, we find ${\varepsilon(w)}/{\varepsilon(r)}=e^{2(\Delta G_w-\Delta G_r)}$. In this limit, the ratio of incorrect monomers to correct ones involves the square of the binding free energy difference, reflecting the fact that two steps of discrimination have occurred.
 
We may also write down expressions for the expected chemical work done per net step of the process. This quantity will involve the total current to absorbing states of the step-wise process for starting with a copolymer $\&xy$, which we may write as:
\begin{eqnarray}
\nonumber
J_{\rm Tot}(y,x)&=&\frac{1}{\mathcal{N}(y,x)}(\Lambda_1^+(r,y)Q(w,y)Q(y,x)\\\nonumber&+&\Lambda_1^+(w,y)Q(r,y)Q(y,x)+\Lambda_1^-(y,x)Q(r,y)Q(w,y)),\\
\label{eq:HopfCurrent}
\end{eqnarray}
where $\mathcal{N}$ is a normalisation factor that will cancel out of calculations. In order to track each of the terms here, we shall break down the contributions to the chemical work done into three parts, one for each of the petals present in the step-wise process. These three petals correspond to adding a monomer type $r$, adding a monomer type $w$ or removing a monomer type $y$. Let us label each of these contributions to the chemical work with a subscript, $\mathcal{G}_r(y,x)$ for the transition $\&xy\to\&xyr$, $ \mathcal{G}_w(y,x)$ for the transition $\&xy\to\&xyw$, and $ \mathcal{G}_q(y,x)$ for the transition $\&xy\to\&x$. From the $r$ petal, we have:
\onecolumngrid
\begin{eqnarray}
\nonumber
 \mathcal{G}_r(y,x)&=&\Bigg[\left(-\Delta G_{\act}+\ln\frac{M_{in}}{M_{\act}}\right)k_1k_{\act}k_{KP}e^{-\Delta G_r}(M_{in}+M_{\act}e^{\Delta G_{\act}})\\\nonumber
&+&(\Gpol+\Delta G_r-\Delta G_y+\ln M_{in}-\Delta G_a)(k_1k_{\act}k_{\pol}M_{in}e^{-\Delta G_y})\\\nonumber
&+&(\Gpol+\Delta G_r-\Delta G_y+\ln M_{\act})k_{KP}k_{\pol}M_{\act}e^{-\Delta G_y}(k_1e^{-\Delta G_r}+k_{\act})\Bigg]\\
&\times&\frac{Q(w,y)Q(y,x)}{\mathcal{N}(y,x)J_{Tot}(y,x)}.
\label{eq:Fryx}
\end{eqnarray}
\twocolumngrid
The first line of eqn.~\ref{eq:Fryx} corresponds to the chemical work associated with the internal cycle (inactive monomer binds, gets activated, and activated monomer unbinds). The second line corresponds to an external cycle: an inactive monomer binds to the template, is activated and is polymerised into the chain with the previous monomer, $y$, detaching from the template. The third line corresponds to the alternative external cycle: an active monomer binds to the template and is polymerised with monomer $y$ unbinding from the template. We may similarly write down $ \mathcal{G}_w(y,x)$ as eqn.~\ref{eq:Fryx}, except swapping $r$ and $w$. Finally, the contribution to the chemical work from the petal for removing monomer $y$ may be written:
\onecolumngrid
\begin{eqnarray}
\nonumber
 \mathcal{G}_q(y,x)&=&\Bigg[-(\Gpol+\Delta G_y-\Delta G_x+\ln M_{in}-\Delta G_a)(k_1k_{\act}k_{\pol}e^{\Delta G_{\act}-\Delta G_y-\Gpol})\\\nonumber
&-&(\Gpol+\Delta G_y-\Delta G_x+\ln M_{\act})k_{KP}k_{\pol}e^{-\Gpol-\Delta G_y}(k_1e^{-\Delta G_r}+k_{\act})\Bigg]\\
&\times&\frac{Q(r,y)Q(w,y)}{\mathcal{N}(y,x)J_{Tot}(y,x)}.
\end{eqnarray}
\twocolumngrid
Here, only external cycles are possible. The first line corresponds to monomer $x$ rebinding to the template, monomer $y$ being depolymerised, this monomer being deactivated and an inactive monomer $y$ unbinding from the template; and the second line to $x$ rebinding, $y$ being depolymerised and active monomer $y$ unbinding from the template. The distribution, $\xi(y,x)$ may be calculated from eqn.~\ref{eq:xi} and $P$ from eqn.~\ref{eq:P} (both demonstrated in the supporting information), letting the chemical work done per net step of the the 1-loop model be written:
\begin{eqnarray}
\nonumber
\Delta  \mathcal{G}= \frac{1}{2P-1}\sum_{x,y\in\{r,w\}}\xi(y,x)\left( \mathcal{G}_r(y,x)+ \mathcal{G}_w(y,x)+ \mathcal{G}_q(y,x)\right).\\
\end{eqnarray}

This chemical work done is plotted for a certain set of parameters in figure~\ref{fig:HopfData}\,(b) and is also compared both to the results of direct simulation and the simpler ``0-loop" model which has chemical work, $\Gpol$. The free-energy cost of the proofreading mechanism diverges as $\Delta G_{\pol}\to\Gamma$ since there will be a finite chemical work done per monomer addition/removal step due to the proofreading internal cycle, and the number of addition/removal steps per net step diverges. Further, for large $\Delta G_{\pol}$, the work tends to be dominated by $\Delta G_{\pol}$, albeit very slowly, as shown by the orange line gradually approaching $\Gpol$ (the blue line) in figure~\ref{fig:HopfData}(b). 
 
Additionally, we can find an expression for the time taken per net step forwards, eqn.~\ref{eq:ExpectedTime}. For this quantity, we need  the explicit expression for the normalisation, $\mathcal{N}$. Similarly to the chemical work, we can split this term into contributions from the petal adding an $r$, $\mathcal{N}_r(y,x)$; from the petal adding a $w$, $\mathcal{N}_w(y,x)$; from the petal removing monomer $y$, $\mathcal{N}_q(y,x)$ and a contribution from the central node. These normalisation terms come from the sums of spanning trees directed to the individual nodes in the closed step-wise process. We see that
\begin{eqnarray}
\nonumber \mathcal{N}_r(y,x)&=&\Big[k_1M_{\text{in}}(k_{\act}e^{\Delta G_{\text{act}}}+k_{KP}e^{-\Delta G_r}+k_{\pol}e^{-\Delta G_y})\\\nonumber+k_{KP}k_{\act}&M_{\text{act}}&e^{\Delta G_{\text{act}}}+k_1k_{\act}M_{\text{in}}+k_{KP}k_{\act}M_{\text{act}}+k_1k_{KP}e^{-\Delta G_r}\Big]\\
&\times&Q(y,x)Q(w,y),
\end{eqnarray}
with a similar result for $\mathcal{N}_w(y,x)$ except swapping $r$ and $w$. Finally, for the monomer removal petal, we have:
\begin{eqnarray}
\nonumber
\mathcal{N}_q(y,x)&=&k_{\pol}e^{-\Gpol}(k_1e^{-\Delta G_y}+k_{\act}\\
&+&k_{\act}e^{\Delta G_{\text{act}}})Q(r,y)Q(w,y).
\end{eqnarray}
The total normalisation is then:
\begin{eqnarray}
\nonumber
\mathcal{N}(y,x)&=&\mathcal{N}_r(y,x)+\mathcal{N}_w(y,x)+\mathcal{N}_q(y,x)\\
&+&Q(y,x)Q(r,y)Q(w,y),
\end{eqnarray}
with the last term being the contribution from the starting, central node. This normalisation can be used in eqn.~\ref{eq:HopfCurrent} to give the current to absorbing states, which can be used in eqn.~\ref{eq:ExpectedTime} to find the expected time per net step. This time is plotted in figure~\ref{fig:HopfData}\,(c), alongside a simulation of the same model and the simplified 0-loop model for comparison. Like the chemical work in figure~\ref{fig:HopfData}\,(b), the time per net step diverges as $\Gpol\to\Gamma$, since each monomer addition/removal step will take finite time, but the number of such steps required for a net forwards step diverges. Unsurprisingly, the time taken for a given driving for the Hopfield model is longer than that of the simple model, due to the proofreading cycle. 
 
Hopfield's model for proofreading may be naturally extended to include $N$ activation stages instead of just one.\citeNHop 
We shall call these extensions the $N$-loop Hopfield models. These models can be solved recursively to write down expressions for the sums over spanning trees, $\Lambda^\pm_N(y,x),\;Q_N(y,x)$, as a function of the number of loops, $N$. We shall consider the model as in figure~\ref{fig:HopfModel}\,(b). A detailed derivation of the sums over spanning trees is given in appendix~\ref{App:NHopfEqns}. From these sums over spanning trees, we calculate the bulk frequencies, the time taken per net step and the chemical work done per net step using recursive relations (see appendix~\ref{App:NHopfEqns}). 

For simplicity, we shall discuss the case where the monomer binding free energy is only dependent on monomer type, not on activation stage; each activation stage is associated with a free energy change of $\Delta G_{\act}$; each active monomer is present in the environment at a concentration $M_{\act}$ except the inactive monomers at concentration $M_{\ina}$; and the overall rate constants are $k_1$ for binding of inactive monomer, $k_{\KP}$ for binding of active monomers, $k_{\act}$ for activation of monomers. Under these assumptions, the corresponding rates are given in appendix~\ref{App:NHopfEqns}.

\begin{figure}
    \centering
    \begin{subfigure}{0.48\textwidth}
        \includegraphics[scale=0.5]{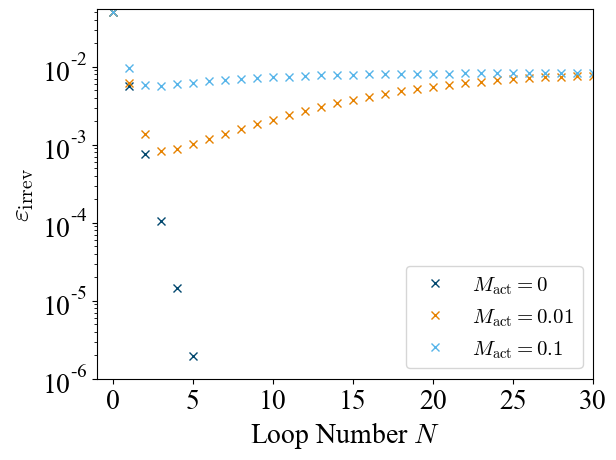}
    \end{subfigure}
    \begin{subfigure}{0.48\textwidth}
        \includegraphics[scale=0.5]{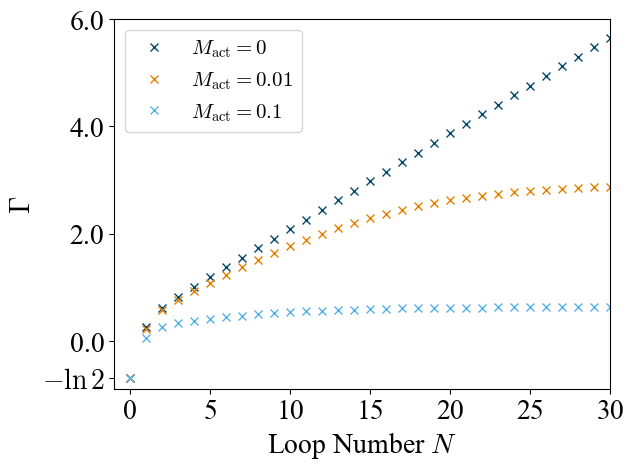}
    \end{subfigure}

    \caption{ Many-loop models have limited efficacy for finite $M_{\act}$ for the proofreading model introduced in figure~\ref{fig:HopfModel}. Plots of the error in the irreversible limit, $\varepsilon_{\irr}(w)$ and the stall point driving, $\Gamma$, for different values of the active monomer concentrations, $M_{\act}$, and other parameters: $\Delta G_r=2,\;\Delta G_w=-2,\;M_{\ina}=1,\;\Delta G_{\act}=-1,\;ks=1$. The $N=0$, error is not shown for clarity, but is $0.5$ for all $M_{\act}$.}
    \label{fig:NLoopGamErr}
\end{figure}

To reduce the frequency of incorrect monomers in the product, we wish to have a low concentration $M_{\act}$ of active monomers in solution to force the system into utilising the proofreading cycles. Indeed, the bulk error probability in the irreversible limit (calculated using eqn.~\ref{eq:IrreversibleError} and plotted in figure~\ref{fig:NLoopGamErr}\,(a) shows a strong improvement with loop number for low $M_{\act}$, but larger values of $M_{\act}$ lead to much worse performance and limited (or negative) returns to increasing the number of loops. 

However, for finite driving strength $\Delta G_{\rm pol}$, we cannot allow this concentration to be arbitrarily small. To see why, consider the stall point, $\Gamma(N)$, derived in appendix~\ref{App:NHopfEqns} and plotted for a certain set of parameters in figure~\ref{fig:NLoopGamErr}\,(b). It is observed that the stall point driving increases monotonically with $N$, and that this increase is faster and tends to a higher limit for smaller $M_{\act}$. We find that the limiting $\Gamma$ scales approximately linearly with $-\ln(M_{\act})$. Intuitively, introducing more monomer states at low concentration in the environment destabilises the polymer. For small $M_{\act}$ and driving $\Delta G_{\rm pol}$, the depolymerisation of the polymer into these activated states competes with its tendency to grow by binding to and activating the inactive monomers.

One drawback of proofreading with a large number of loops is therefore that the tendency to disassemble the growing polymer increases. A second effect is a tendency to introduce errors by alternate pathways if $M_{\act}$ is non-zero. Specifically, for $M_{\act}\neq0$, we observe in figure~\ref{fig:NLoopGamErr}\,(a) a minimum in $\epsilon_{\rm irrev}(w)$ for a relatively small value of $N$. This minimum can be explained by splitting the pathways by which a monomer can go from solution to being incorporated into the polymer into two, either starting from a fully inactive monomer or from a partially activated one. The pathway starting with an inactive monomer will have the highest discrimination between right and wrong monomers and will improve exponentially with more loops, as demonstrated by the exponential decrease in error for $M_{\act}=0$. However, the probability that a monomer, taking this path, will reach polymerisation falls exponentially with loop number at the same time. On the other hand, the pathway from partially active monomers will give an error that reaches some non-zero limit as the number of loops, $N$, increases. Further, the rate with which activated monomers bind to an available template site and subsequently get incorporated into the polymer will also tend to a constant.  As such, the error will initially decrease exponentially with $N$, but for non-zero $M_{\rm act}$ will eventually become dominated by the less discriminating, partially active monomer pathways through which monomers are more likely to be incorporated into the polymer.

\begin{figure}
    \centering
    \begin{subfigure}{0.48\textwidth}
    \includegraphics[scale=0.5]{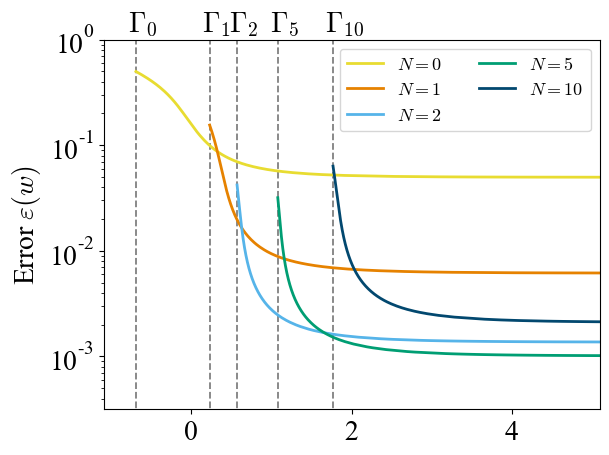}
    \label{fig:NHopfError}
    \end{subfigure}
    \begin{subfigure}{0.48\textwidth}
    \includegraphics[scale=0.5]{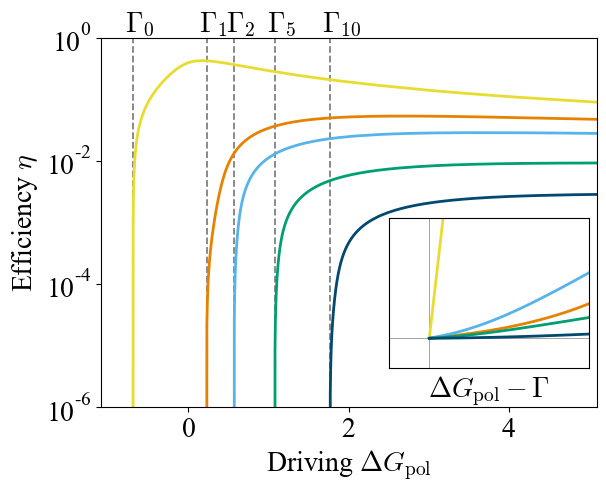}
    \label{fig:NHopfEff}
    \end{subfigure}
    \caption{Plots of (a) the error and (b) efficiency of the $N$-loop proofreading model (figure~\ref{fig:HopfModel}) for a range of $N$, with the same parameters as in the one loop Hopfield case, figure~\ref{fig:HopfData}, as a function of driving $\Delta G_{\rm pol}$. Proofreading is observed to generally increase accuracy above its stall point, but in a thermodynamically inefficient way.  The enhanced plot in the second graph shows the efficiencies near the stall point for each of the loop numbers on a non-logarithmic scale to emphasise the decreasing gradient at stall.}
    \label{fig:NLoopEfficiency}
\end{figure}

Having calculated the error probability $\epsilon(w)$ at finite driving, plotted in figure~\ref{fig:NLoopEfficiency}\,(a); used $\epsilon(x,y)$ to calculate the entropy rate; and calculated $\Delta \mathcal{G}$; we can evaluate the efficiency $\eta$, as in eqn.~\ref{eq:efficiency} (see supporting information for demonstrations). This efficiency is plotted in figure~\ref{fig:NLoopEfficiency} for $N=0,1,5,10$ and a certain set of parameters. Although accuracy is generally increased above the stall point, we see that in this particular model kinetic proofreading requires much more work than the minimum required to generate information and as such are inefficient. Additionally, the gradient of the efficiency at minimum driving, $\Gamma_N$, is zero for $N>0$, reflecting how at minimum driving, the number of monomer addition/removal steps diverges, but the chemical work done per such step remains finite.  

\section{Conclusion}
We have presented a method for analysing copolymerisation models with complex networks of reactions leading to the incorporation or removal of monomers. By coarse graining, a model may be transformed into a simpler model which may be solved and then afterwards, information from the fine-grained process may be put back into the model to extract thermodynamic or kinetic quantities such as chemical work done, molecule exchange or time taken. The approach allows for complex incorporation motifs to be considered alongside nearest neighbour interactions in a thermodynamically well-defined model of polymerisation with microscopic reversibility. We note that all of these features were present in the kinetic proofreading example in Section~\ref{sec:KP}. Moreover, phenomena such as the shift in stall point with loop number and the non-monotonicity of error rate with loop number rely on these features being present in the model.

In general, this method provides a way to extract model predictions numerically quickly and without the need for simulations. Doing so is particularly useful when simulating polymer growth is slow, either due to the details of the incorporation process or because the polymer is near its stall point. Additionally, the approach makes screening of a large parameter space for a given model topology feasible.

In addition to the numerical performance, the approach allows for analytic results in simpler models or those with helpful symmetries, as well as in certain limits for more complex models. The process of summing over spanning trees is particularly well suited to identifying the structure of the process and providing simplified results.

Moving forwards, it is an open question as to whether components of the techniques developed here can be applied outside of the context of infinitely long polymers whose tips have reached steady state. An obvious goal would be a simplified way to analyse finite-length ``oligomers".\cite{poulton2021edge}. More generally, we believe the key equation of this paper, eqn.~\ref{eq:effectiverates}, may be applied more generally for the coarse graining of Markov processes. Specifically, that if a set of states are enclosed between two boundary states, in the sense that any path from one of the trapped states to outside must pass through one of the boundary states, then this set of states may be replaced by a pair of edges analogously to eqn.~\ref{eq:effectiverates} which shall preserve steady state properties of the Markov process.

This framework could be applied to explore models of copolymerisation processes such as those presented in \citeSolvable more straightforwardly or more thoroughly. Alternatively, the method would allow for more complex reaction steps to be included in such models. The framework presented here is particularly useful when backwards steps are relevant, either when the system is weakly driven and thus operating near stall, or when thermodynamics is of importance or interest. We also predict that it will be useful to guide design principles for synthetic copolymerisation systems, which are often particularly well-described by the class of models studied here.

\section*{Supplementary Material}
The supplementary material contains a C++ script implementing the Gillespie algorithm that reproduces the data for the 1-loop Hopfield kinetic proofreading model presented in figure~\ref{fig:HopfData}, and a MATLAB script for numerically calculating quantities of the 1-Loop and N-Loop Hopfield kinetic proofreading models presented in section~\ref{sec:KP} and shown in the solid lines of figure~\ref{fig:HopfData}, the points of figure~\ref{fig:NLoopGamErr} and figure~\ref{fig:NLoopEfficiency}.

\section*{Acknowledgements}
This work is part of a project that has received funding from the European Research Council (ERC) under the European Union’s Horizon 2020 research and innovation program (Grant agreement No. 851910). T.E.O. is supported by a Royal Society University Fellowship. J.J. is  supported by a Royal Society PhD studentship.

\section*{Author Declarations}
\subsection*{Conflict of Interest}
The authors have no conflicts to disclose.
\subsection*{Author Contributions}
All authors conceived of the project. B.Q. produced the methodology and analysis and wrote the initial draft. All authors interpreted results and reviewed and edited this paper.

\section*{Data Availability Statement}
The data that support the findings of this study are reproducible from the code openly available in Zenodo at \url{https://doi.org/10.5281/zenodo.7271702}.

\nocite{*}
\bibliography{aipsamp}

\onecolumngrid
\appendix
\newpage
\section{Factorising sums of spanning trees}
\label{App:SAW}
We note here that sums of spanning trees can be factorised in terms of Self-Avoiding Walks (SAWs), a result which is both useful for generating sets of spanning trees and allows us to make statements about ratios of propensities of balanced models. For a given process, $\mathcal{G}=(\mathcal{X},K)$, for which we wish to find the sum of spanning trees rooted at $x_1\in\mathcal{X}$, we may factorise this sum in terms of self-avoiding walks (SAWs) between two vertices in the graph. Select some other arbitrary vertex $x_2\in\mathcal{X}/\{x_1\}$ and let $\mathcal{S}(x_2,x_1)$ be the set of SAWs from $x_2$ to $x_1$. For each $S\in\mathcal{S}(x_2,x_1)$, we can construct $\mathcal{G}_S=(\{s\}\cup(\mathcal{X}/S),K_S)$ analogously to eqn.~\ref{eq:cycleprocess}, whereby we collapse the nodes in the SAW, $S$, into the single node $s$. The sum over spanning trees rooted at $x_1$ may then be written:
\begin{eqnarray}
\nonumber
    \sum_{T\in\mathcal{T}(x_1)}\prod_{e\in T}K(e)=\sum_{S\in\mathcal{S}(x_2,x_1)}\underbrace{\left[\prod_{e\in S}K(e)\right]}_{\text{SAW term}}\underbrace{\left[\sum_{T\in\mathcal{T}_S(s)}\prod_{e\in T}K_S(e)\right]}_{\text{Spanning tree term}},\\
    \label{eq:SAW-ST}
\end{eqnarray}
where $\mathcal{T}(x),\mathcal{T}_S(x)$ are the sets of spanning trees directed to $x$ for the original process, $\mathcal{G}$, and the new process, $\mathcal{G}_S$. For example, in figure~\ref{fig:SpanTree}(a), the spanning trees are arranged in terms of SAWs from node $1$ to node $3$, with the first row for SAW: $1\to2\to4\to3$; the second row for $1\to2\to3$; and the last three rows for $1\to3$. Similarly for figure~\ref{fig:SpanTree}(b), the trees are arranged in terms of SAWs from node $1$ to node $4$ with row one for $1\to2\to3\to4$; row two for $1\to2\to4$; row three for $1\to3\to4$; and row four for $1\to3\to2\to4$.

\section{Normalisation constant for example absorbing Markov process}
\label{App:ExNorm}
The normalisation constant for the closed example process, figure~\ref{fig:example}(b), can be found by considering the spanning trees rooted at each of the nodes. Factorising these in terms of SAWs, we write:
\begin{eqnarray}
\nonumber
        \mathcal{N}&=&\big[r_{34}r_{42}r_{21}+r_{32}r_{24}k_B+r_{32}r_{21}(r_{43}+r_{42}+k_B)+r_{34}k_B(r_{24}+r_{23}+r_{21})\\\nonumber
        &+&(r_{31}+k_A)(r_{21}r_{43}+r_{21}k_B+r_{42}r_{21}+r_{23}r_{43}+r_{23}k_B+r_{42}r_{23}+r_{24}r_{43}+r_{24k_B})\big]\\\nonumber
        &+&\big[r_{13}r_{34}r_{42}+r_{13}r_{32}(r_{43}+r_{42}+k_B)\\\nonumber
        &+&r_{12}(r_{34}r_{42}+r_{43}r_{32}+r_{32}k_B+r_{32}r_{42}+r_{43}(r_{31}+k_A)+k_B(r_{31}+k_A)+r_{42}(r_{31}+k_A)+r_{34}k_B)
        \big]\\\nonumber
        &+&\big[r_{12}r_{24}r_{43}+r_{12}r_{23}(r_{42}+r_{43}+k_B)
        +r_{13}((r_{43}+k_B)(r_{21}+r_{23}+r_{24})+r_{42}(r_{21}+r_{23}))\big]\\
        &+&\big[r_{13}r_{34}(r_{21}+r_{23}+r_{24})+r_{13}r_{32}r_{24}+r_{12}r_{23}r_{34}+r_{12}r_{24}(r_{31}+k_A+r_{32}+r_{34})\big].
\end{eqnarray}
The first square bracket corresponds to the trees rooted at node $1$, organised by SAWs from node $3$; the second to trees rooted at $2$ organised by SAWs from $1$; the third to trees rooted at $3$ organised by SAWs from $1$ and the fourth to trees rooted at $4$ organised by SAWs from $1$.

\section{Equivalence between chemical work calculated from Edges and cycles.}
\label{App:Edge-Cycle}
Here, we shall show the equivalence of chemical work for a process calculated by summing over edges versus summing over cycles. For this, consider a process $(\mathcal{X},K)$, without any absorbing states (for simplicity) and such that every edge is microscopically reversible and let $\pi(x)$ be the steady state probability to be in state $x$. For an edge $x\leftrightharpoons y$, as described in section~\ref{sec:absorbingwork}, the net current through this edge is:
\begin{eqnarray}
        J_{x\leftrightharpoons y}=\pi(x)K(x,y)-\pi(y)K(y,x).
\end{eqnarray}
We can write $\pi(x)$ in terms of spanning tress by MCTT, and by appendix~\ref{App:SAW}, we may expand the sum over spanning trees by SAWs from $y$ to $x$. For $\pi(y)$, we may expand by SAWs from $x$ to $y$ such the spanning tree terms of both expansions are the same and only the direction of edges in the SAW terms is flipped. The net current may then be written:
\begin{eqnarray}
        J_{x\leftrightharpoons y}=\frac{1}{\mathcal{N}}\sum_{S\in\mathcal{S}(y,x)}\left[K(x,y)\prod_{e\in S}K(e)-K(y,x)\prod_{e'\in S}K(e')\right]\left[\sum_{T\in\mathcal{T}_S(s)}\prod_{e\in T}K_S(e)\right]
\end{eqnarray}
where $\mathcal{S}(x,y)$ is the set of SAWs from node $x$ to node $y$; $\mathcal{N}$ is the normalisation as in eqn.~\ref{eqn:MCTTss}, and $e'$ is the edge in the opposite direction, i.e. if $e=x\to y$, $e'=y\to x$; and the last bracketed term is the spanning tree part for SAW, $S$, as in eqn.~\ref{eq:SAW-ST}. One of the SAWs from $y$ to $x$ will simply be the single transition $x\to y$, however, this term will cancel out from the sum leaving just the non-trivial SAWs. Taking a non-trivial SAW from $y$ to $x$ and multiplying by the rate $K(x,y)$ gives a cycle containing the edge $x\to y$. Therefore, the current may be written as a sum over cycle currents, as in section~\ref{sec:absorbingwork}, of cycles which contain the edge $x\to y$ minus those which contain $y\to x$. Each of the edges contains a contribution to chemical work $\ln\left(\frac{K(x,y)}{K(y,x)}\right)$. The total chemical work before absorption is the sum over all edges of these contributions:
\begin{eqnarray}
        \mathcal{W}_{\rm chem}=\sum_{x\leftrightharpoons y}\ln\left(\frac{K(x,y)}{K(y,x)}\right)\frac{J_{x\leftrightharpoons y}}{J_{\rm Tot}}.
\end{eqnarray} 
Since, in this sum the $J_{x\leftrightharpoons y}$ may be split up as a sum over cycles, we may collect the parts of this corresponding to given cycles and convert the sum over edges into a sum over cycles. Doing so we find the contribution to the chemical work from cycle, $C$, to be $ln\left(\frac{A(C)}{A(C')}\right)$, i.e. the affinities as we might expect. Hence, the sum over cycles is equivalent to the sum over edges.

\section{Cycles of the example absorbing process}
\label{App:CycleList}
We make divide the cycles of the example process, figure~\ref{fig:example}(a), into internal cycles, external cycles to absorbing state $A$ and external cycles to absorbing state $B$. Firstly, the internal cycles are:
\begin{center}
\begin{tabular}{|c|c|c|}
    \hline
    \tikz[baseline=5ex]{
    \tikzstyle{vertex}=[circle,fill=black,inner sep=0pt,minimum size=4pt]
    \tikzstyle{edge}=[-, thick]
    \node[vertex, label=1] (1) at (0,0) {};
    \node[vertex, label=3] (3) at (12ex,0) {};
    \node[vertex, label=2] (2) at (6ex,10.4ex) {};
    \node[vertex, label=4] (4) at (18ex,10.4ex) {}; 
    \draw[edge](1)--(2);
    \draw[edge](2)--(4);
    \draw[edge](4)--(3); 
    \draw[edge](3)--(1);
    }
    &
    \tikz[baseline=5ex]{
    \tikzstyle{vertex}=[circle,fill=black,inner sep=0pt,minimum size=4pt]
    \tikzstyle{edge}=[-, thick]
    \node[vertex, label=1] (1) at (0,0) {};
    \node[vertex, label=3] (3) at (12ex,0) {};
    \node[vertex, label=2] (2) at (6ex,10.4ex) {};
    \draw[edge](1)--(2);
    \draw[edge](2)--(3);
    \draw[edge](3)--(1);
    }
    &
   \tikz[baseline=5ex]{
    \tikzstyle{vertex}=[circle,fill=black,inner sep=0pt,minimum size=4pt]
    \tikzstyle{edge}=[-, thick]
    \node[vertex, label=3] (3) at (12ex,0) {};
    \node[vertex, label=2] (2) at (6ex,10.4ex) {};
    \node[vertex, label=4] (4) at (18ex,10.4ex) {}; 

    \draw[edge](2)--(3);
    \draw[edge](3)--(4); 
    \draw[edge](4)--(2);
    }
    \\&&\\
    $1\to2\to4\to3\to1$
    &
    $1\to2\to3\to1$
    &
    $2\to4\to3\to2$
    \\\hline
\end{tabular} 
\end{center}
where the cycle is written out below in the clockwise direction. Similarly, we find the external cycles to state $A$:

\begin{center}
\begin{tabular}{|c|c|c|}
    \hline
    \tikz[baseline=5ex]{
    \tikzstyle{vertex}=[circle,fill=black,inner sep=0pt,minimum size=4pt]
    \tikzstyle{edge}=[->, thick]
    \node[vertex, label=1] (1) at (0,0) {};
    \node[vertex, label=3] (3) at (12ex,0) {};
    \node[vertex, label=A] (A) at (24ex,0) {};
    \draw[edge](1)--(3);
    \draw[edge](3)--(A);
    }
    &
    \tikz[baseline=5ex]{
    \tikzstyle{vertex}=[circle,fill=black,inner sep=0pt,minimum size=4pt]
    \tikzstyle{edge}=[->, thick]
    \node[vertex, label=1] (1) at (0,0) {};
    \node[vertex, label=3] (3) at (12ex,0) {};
    \node[vertex, label=2] (2) at (6ex,10.4ex) {};
    \node[vertex, label=A] (A) at (24ex,0) {};
    \draw[edge](1)--(2);
    \draw[edge](2)--(3);
    \draw[edge](3)--(A);
    }
    &
   \tikz[baseline=5ex]{
    \tikzstyle{vertex}=[circle,fill=black,inner sep=0pt,minimum size=4pt]
    \tikzstyle{edge}=[->, thick]
    \node[vertex, label=1] (1) at (0,0) {};
    \node[vertex, label=3] (3) at (12ex,0) {};
    \node[vertex, label=2] (2) at (6ex,10.4ex) {};
    \node[vertex, label=4] (4) at (18ex,10.4ex) {};
    \node[vertex, label=A] (A) at (24ex,0) {};

    \draw[edge](1)--(2);
    \draw[edge](2)--(4); 
    \draw[edge](4)--(3);
    \draw[edge](3)--(A);
    }
    \\&&\\
    $1\to3\to A$
    &
    $1\to2\to3\to A$
    &
    $1\to2\to4\to3\to A$
    \\\hline
\end{tabular} 
\end{center}
Finally, the external cycles to absorbing state $B$ are:
\begin{center}
\begin{tabular}{|c|c|c|c|}
    \hline
    \tikz[baseline=5ex]{
    \tikzstyle{vertex}=[circle,fill=black,inner sep=0pt,minimum size=4pt]
    \tikzstyle{edge}=[->, thick]
    \node[vertex, label=1] (1) at (0,0) {};
    \node[vertex, label=2] (2) at (6ex,10.4ex) {};
    \node[vertex, label=4] (4) at (18ex,10.4ex) {};
    \node[vertex, label=B] (B) at (24ex,10.4ex) {};
    \draw[edge](1)--(2);
    \draw[edge](2)--(4);
    \draw[edge](4)--(B);
    }
    &
    \tikz[baseline=5ex]{
    \tikzstyle{vertex}=[circle,fill=black,inner sep=0pt,minimum size=4pt]
    \tikzstyle{edge}=[->, thick]
    \node[vertex, label=1] (1) at (0,0) {};
    \node[vertex, label=3] (3) at (12ex,0) {};
    \node[vertex, label=2] (2) at (6ex,10.4ex) {};
    \node[vertex, label=4] (4) at (18ex,10.4ex) {};
    \node[vertex, label=B] (B) at (24ex,10.4ex) {};
    \draw[edge](1)--(2);
    \draw[edge](2)--(3);
    \draw[edge](3)--(4);
    \draw[edge](4)--(B);
    }
    &
   \tikz[baseline=5ex]{
    \tikzstyle{vertex}=[circle,fill=black,inner sep=0pt,minimum size=4pt]
    \tikzstyle{edge}=[->, thick]
    \node[vertex, label=1] (1) at (0,0) {};
    \node[vertex, label=3] (3) at (12ex,0) {};
    \node[vertex, label=2] (2) at (6ex,10.4ex) {};
    \node[vertex, label=4] (4) at (18ex,10.4ex) {};
    \node[vertex, label=B] (B) at (24ex,10.4ex) {};

    \draw[edge](1)--(3);
    \draw[edge](3)--(2); 
    \draw[edge](2)--(4);
    \draw[edge](4)--(B);
    }
    &
    \tikz[baseline=5ex]{
    \tikzstyle{vertex}=[circle,fill=black,inner sep=0pt,minimum size=4pt]
    \tikzstyle{edge}=[->, thick]
    \node[vertex, label=1] (1) at (0,0) {};
    \node[vertex, label=3] (3) at (12ex,0) {};
    \node[vertex, label=4] (4) at (18ex,10.4ex) {};
    \node[vertex, label=B] (B) at (24ex,10.4ex) {};
    \draw[edge](1)--(3);
    \draw[edge](3)--(4);
    \draw[edge](4)--(B);
    }
    \\&&&\\
    $1\to2\to4\to B$
    &
    $1\to2\to3\to4\to B$
    &
    $1\to3\to1\to4\to B$
    &
    $1\to3\to4\to B$
    \\\hline
\end{tabular} 
\end{center}
\section{Number of steps per net forward step of a random walk}
\label{App:2P-1}
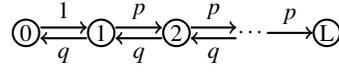
\begin{figure}[ht!]
    \centering
    \begin{tikzpicture}
    \tikzstyle{vertex}=[circle,draw=black, thick,inner sep=0pt,minimum size=10pt]
    \tikzstyle{edge}=[->, thick]
    \node[vertex] (0) at (0,0) {0};
    \node[vertex] (1) at (1,0) {1};
    \node[vertex] (2) at (2,0) {2};
    \node[circle,inner sep=0pt,minimum size=10pt] (3) at (3,0) {$\cdots$};
    \node[vertex] (L) at (4,0) {L};

    \draw[transform canvas={yshift=0.5ex},edge](0)--node[above]{$1$}(1);
    \draw[transform canvas={yshift=-0.5ex},edge](1)--node[below]{$q$}(0);
    \draw[transform canvas={yshift=0.5ex},edge](1)--node[above]{$p$}(2);
    \draw[transform canvas={yshift=-0.5ex},edge](2)--node[below]{$q$}(1);
    \draw[transform canvas={yshift=0.5ex},edge](2)--node[above]{$p$}(3);
    \draw[transform canvas={yshift=-0.5ex},edge](3)--node[below]{$q$}(2);
    \draw[edge](3)--node[above]{$p$}(L);
    
    \end{tikzpicture}
    \caption{Graphical representation of the random walk process considered.}
    \label{fig:RWgraph}
\end{figure}

Here we shall derive the number of steps per net forward step of a random walk. Let us set up a random walk as follows. Let the state space be the nodes $\{0,1,\cdots L\}$ where $L$ is the length of the walk (polymer). Let the transition $0\to1$ have probability $1$, $i\to i+1$ for $i=1,\cdots L-1$ have probability $p$,  $i\to i-1$ for $i=1,\cdots L-1$ have probability $q=1-p$ and let state $L$ be an absorbing state as in figure~\ref{fig:RWgraph}.

We then wish to find the expected number of steps to absorption, given starting in state $0$, for which we can utilise the spanning tree methods with eqn.~\ref{eq:ExpectedTime}. Since the total rate out of any state sums to one, the expected number of steps equals the expected time to absorption. Thus, we can form the closed process starting at $0$. Let $f(n)$ be the sum over spanning trees rooted at node $n$ for the closed process. $f(n)$ is given by:
\begin{eqnarray}
        f(n)=
        \begin{cases}
        \sum\limits_{i=0}^{L-1}p^iq^{L-1-i}\;&\text{for}\;n=0\\
        p^{n-1}\sum\limits_{i=0}^{L-1-n}p^iq^{L-1-n-i}\;&\text{for}\;n=1,\cdots, L-1
        \end{cases}.
\end{eqnarray}
From this, the expected number of steps before absorption is:
\begin{eqnarray}
        \mathbb{E}[\text{steps}]=\frac{\sum\limits_{n=0}^{L-1}f(n)}{pf(L-1)}.
\end{eqnarray}
By utilising the formulae for finite geometric series, we can find the expected number of steps to be:
\begin{eqnarray}
        \mathbb{E}[\text{steps}]=\frac{1}{2p-1}\left(L-1-\frac{q}{p^L}\left(\frac{p^L-q^L}{p-q}\right)+\frac{q^L}{p^L}+\frac{p^L-q^L}{p^{L-1}}\right).
\end{eqnarray}
Most of this expression is sub-linear in $L$, and as such:
\begin{eqnarray}
        \lim_{L\to\infty}\frac{\mathbb{E}[\text{steps}]}{L}=\frac{1}{2p-1},
\end{eqnarray}
which is the net number of steps per net forward step.

\section{The frequency at stall is given by the diagonal cofactors of a matrix}
\label{App:Estall}
We wish to show that, at stall, the frequency with which a monomer appears in the bulk of the copolymer is proportional to the cofactor of the corresponding diagonal element of a matrix: 
\begin{eqnarray}
        \varepsilon(x)\propto A_{xx},
\end{eqnarray}
where $A_{ij}$ is the cofactor of element $i,j$ of the matrix $\mathbbm{1}-Z$. To show this relation we will rely on the relationship between cofactors and vectors of the nullspace of a matrix. Let $M$ be an arbitrary matrix with a one dimensional nullspace, and let $A$ be its matrix of cofactors. Recall that
\begin{eqnarray}
        M A^T=\det(M)\mathbbm{1}=0.
\end{eqnarray}
Thus, any column of $A^T$ is in the nullspace of $M$. In anticipation, let $\overrightarrow{\mu}$ be a vector in the nullspace of $M$ and $\overrightarrow{v}$ be a vector in the nullspace of $M^T$. Since $M$ has a one dimensional nullspace, then 
\begin{eqnarray}
        \frac{\mu_x}{\mu_y}=\frac{A_{ix}}{A_{iy}},
\end{eqnarray}
for some arbitrary $i$. Similarly,
\begin{eqnarray}
        \frac{v_x}{v_y}=\frac{A_{xj}}{A_{yj}},
\end{eqnarray}
for arbitrary $j$. 

Looking at eqns.~\ref{eq:v},~\ref{eq:mu}, noting that near the stall point, $v_z<<\omega_{\pm y,x}$, we see that, the tip probabilities, $\mu(x)$, form a vector in the nullspace of $\mathbbm{1}_M-Z$ and the tip velocities, $v_x$, form a vector in the nullspace of $\mathbbm{1}_M-Z^T$. Hence, we have that
\begin{eqnarray}
        \frac{\mu(y)v_y}{\mu(x)v_x}=\frac{A_{jy}A_{yi}}{A_{jx}A_{xi}},
\end{eqnarray}
for arbitrary $i,j$. Thus, we may choose $j=y$ and $i=x$ leading to cancellation such that
\begin{eqnarray}
        \frac{\mu(x)v_x}{\mu(y)v_y}=\frac{A_{xx}}{A_{yy}}.
\end{eqnarray}
Since
\begin{eqnarray}
        \varepsilon(x)=\frac{\mu(x)v_x}{\sum\limits_y\mu(y)v_y}=\frac{A_{xx}}{\sum\limits_yA_{yy}},
\end{eqnarray}
we get the required result. 

\section{The frequencies in the irreversible limit are given by the steady state of a process of the complete graph}
\label{App:Eirrev}
We wish to find an expression for the frequency with which monomer $x$ appears in the bulk of the copolymer in the irreversible limit. This limit is such that the backwards propensities, $\omega_{-yx}=0$. With this assumption, from eqn.~\ref{eq:v},  we have
\begin{eqnarray}
        v_x=\sum_{y}\omega_{+yx}.
\end{eqnarray}
With this form for the velocities, we may manipulate eqn.~\ref{eq:mu}:
\begin{eqnarray}
        \nonumber
        \mu(x)&=&\sum_{y}\frac{\omega_{+yx}}{\sum\limits_z\omega_{+zx}}\mu(y),\\
        \sum_{z\neq x}\omega_{+zx}\mu(x)+\omega_{+xx}\mu(x)&=&\sum_{y\neq x}\omega_{+xy}\mu(y)+\omega_{+xx}\mu(x).
\end{eqnarray}
This last line is the equation for steady state of a Markov process with probability $\mu(x)$ to be in state $x$ and rate $\omega_{+yx}$ of transition from state $x$ to state $y$. Thus, set $\mu(x)$ to be the steady state probability distribution of the Markov process on $M$ states with transition rates from state $x$ to $y$ given by $\omega_{+yx}$, and $v_x=\sum_{y}\omega_{+yx}$. Then, calculating
\begin{eqnarray}
        \varepsilon(x)\propto\mu(x)v_x,
\end{eqnarray}
gives the required result. Finding the distribution, $\mu(x)$, in terms of spanning trees of the complete graph on $M$ elements gives eqn.~\ref{eq:IrreversibleError}.

\section{Simplification of results for factorisable ratios of propensities}
\label{App:factorisable}
We shall show that, if the ratio of propensities factorises as in eqn.~\ref{eq:FactorisationRates}, then we may simplify the stall condition and frequency of monomers at stall. Thinking of the functions $X$ and $Y$ as column vectors, since they have a discrete domain, the matrix $Z$ may be written,
\begin{eqnarray}
        Z=\overrightarrow{X}\overrightarrow{Y}^T.
\end{eqnarray}
By a well known result\cite{harville1998matrix},
\begin{eqnarray}
        \det(\mathbbm{1}_M-\overrightarrow{X}\overrightarrow{Y}^T)=1-\overrightarrow{Y}^T\overrightarrow{X}=1-\sum_{x}X(x)Y(x).
\end{eqnarray}
rearranging gives eqn.~\ref{eq:FactorisedStallCondition}. At stall, this bound is saturated. As shown, the frequency of monomer $x$ in the bulk of the copolymer is given by the cofactor of the diagonal elements of $\mathbbm{1}_M-\overrightarrow{X}\overrightarrow{Y}^T$. The cofactor, $A_{xx}$, may be written:
\begin{eqnarray}
        A_{xx}=\det(\mathbbm{1}_{M-1}-\overrightarrow{X}_{[x]}\overrightarrow{Y}_{[x]}^T)=1-\sum_{y\neq x}X(y)Y(y)=X(x)Y(x),
\end{eqnarray}
using the stall condition, and where $\overrightarrow{X}_{[x]}$ is the vector $\overrightarrow{X}$, missing element $X(x)$, i.e. ${\overrightarrow{X}_{[x]}=(X(1),\cdots X(x-1),X(x+1),\cdots X(M))^T}$. Additionally, because of the stall condition $\sum_x X(x)Y(x)=1$,
\begin{eqnarray}
        \varepsilon_{\text{stall}}(x)=X(x)Y(x)
\end{eqnarray}
is already normalised.

\section{Frequency for a balanced model with two monomer types in the slow polymerisation limit}
\label{App:SlowPolyDB}
We shall derive the frequency of monomer $x$ in the bulk of the copolymer with propensities given by eqn.~\ref{eq:DBProp}, cancelling $n_{\abs}k_{\abs}$, with $R_{\abs}^-=e^{-\Gpol}$, and with $M=2$. With these propensities eqn.~\ref{eq:v} becomes
\begin{eqnarray}
        v_1&=&\frac{v_1}{e^{-\Gpol}+v_1}+\frac{e^{-DG}v_2}{e^{-\Gpol}+v_2}\\
        v_2&=&\frac{e^{DG}v_1}{e^{-\Gpol}+v_1}+\frac{v_2}{e^{-\Gpol}+v_2},
\end{eqnarray}
where $DG=\Delta G_1-\Delta G_2$. These equations may be solved by the following form the velocities:
\begin{eqnarray}
        v_x=e^{\Delta G_y-\Delta G_x}v_y.
\end{eqnarray}
Doing so, reduces eqn.~\ref{eq:v} to a quadratic equation,
\begin{eqnarray}
     0=e^{DG}v_1^2+(1+e^{DG})(e^{-\Gpol}-1)v_1+e^{-\Gpol}(e^{-\Gpol}-2)
\end{eqnarray} 
with one positive root,
\begin{eqnarray}
        v_1=\frac{1}{2}\left((1-e^{-\Gpol})(e^{-DG}+1)+\sqrt{(e^{-\Delta G_{\pol}}-1)^2(e^{-DG}-1)^2+4e^{-DG}}\right),
\end{eqnarray}
when the system is not stalling. $v_2$ can be found from in terms of $v_1$ as $v_2=e^{DG}v_1$. Further, a quick check confirms $v_1=0$ if $\Gpol=-\ln2$. Further, with $v_y$ known, eqn.~\ref{eq:mu} is a simple linear equation, $\mu$ can be found as the eigenvector of the matrix
\begin{eqnarray}
        \begin{pmatrix}
        \frac{1}{e^{-\Gpol}+v_1} & \frac{e^{DG}}{e^{-\Gpol}+v_1}\\
        \frac{e^{-DG}}{e^{-\Gpol}+v_2} & \frac{1}{e^{-\Gpol}+v_2}
        \end{pmatrix},
\end{eqnarray}
with eigenvalue $1$ and normalised to sum to $1$. Combining the solutions for $\mu$ and $v$, using eqn.~\ref{eq:bulk_prob}, gives eqn.~\ref{eq:DBErr}.

\section{Model used for balanced on-rate vs off-rate discrimination comparisons}
\label{App:DBModels}
\begin{figure}[ht!]
    \centering
    \includegraphics[scale=0.5]{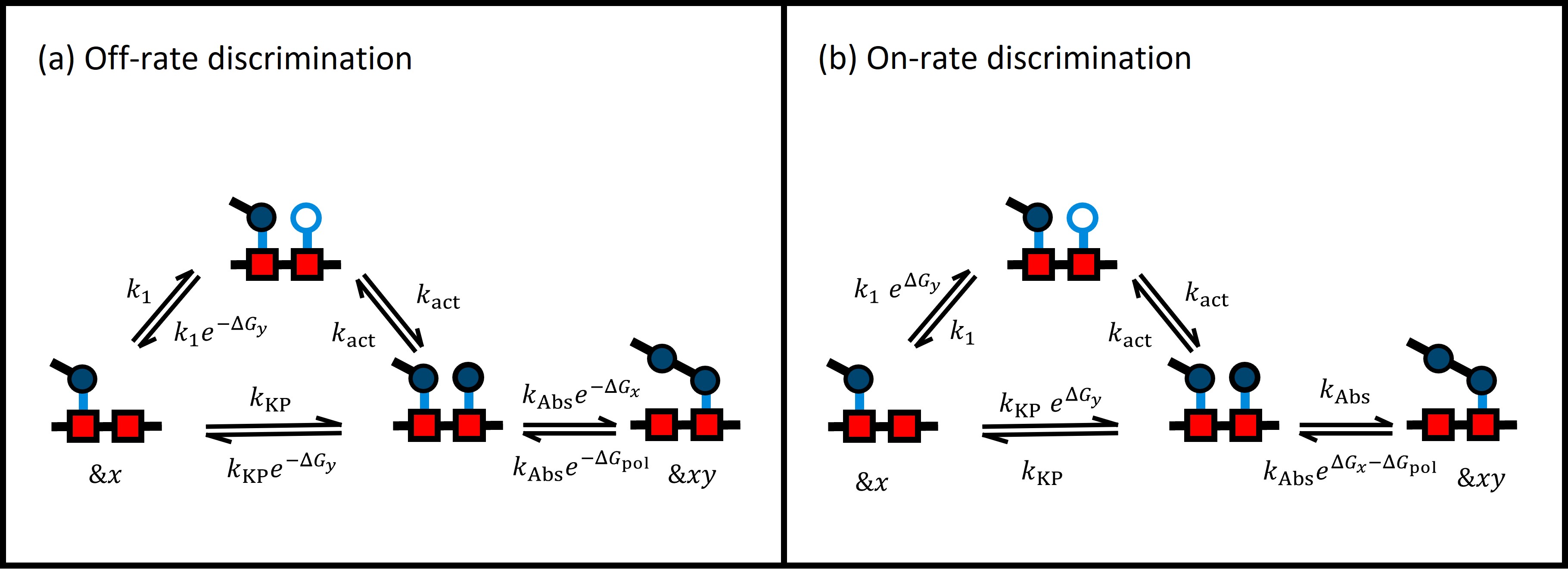}
    \caption{Reaction rates of the (a) off-rate and (b) on-rate discrimination models used to produce the results of figure~\ref{fig:onoffdiscrimination}. These reactions represent a single petal of the step-wise process (figure~\ref{fig:petaldiagram}) between completed states $\&x$ and $\&xy$. For the results in figure~\ref{fig:onoffdiscrimination} for the off-rate and on-rate curves, the following parameters were takes, $\Delta G_1=2,\;\Delta G_2=-2$, $k_1=k_{\KP}=k_{\act}=1$, $k_{\abs}=100$.}
    \label{fig:OnOffReactions}
\end{figure}

\section{Equations for $N$-loop Hopfield model}
\label{App:NHopfEqns}
The sum over spanning trees of the $N$-loop model can be written in terms of sums over spanning trees of the lower loop number models. We label the reaction rates for the $N$-loop process as shown in figure~\ref{fig:HopfModel}. The $N$-loop model has one more node and two more edges than the $N-1$-loop model. Let a subscript, $N$, denote the sums over spanning trees for the $N$-loop models. Tracking the spanning trees, we see,
\begin{eqnarray}
\Lambda^+_N&=&R^+_{\text{act}}(N)\Lambda^+_{N-1}+R^+_{KP}(N)\frac{R^+_{\pol}}{R^-_{\pol}}\sum_{i=0}^N\left[\prod_{j=0}^{i-1}R^+_{\text{act}}(N-j)\right]\Lambda^-_{N-1-i},\\
\Lambda^-_N&=&R^-_{\text{act}}(N)\Lambda^-_{N-1}+R^-_{KP}(N)\sum_{i=0}^N\left[\prod_{j=0}^{i-1}R^+_{\text{act}}(N-j)\right]\Lambda^-_{N-1-i},\\
Q_N&=&\frac{1}{R^-_{\pol}}\left(\Lambda^-_N+R^+_{\pol}\sum_{i=0}^N\left[\prod_{j=0}^{i-1}R^+_{\text{act}}(N-j)\right]\Lambda^-_{N-1-i}\right),
\end{eqnarray}
with initial conditions 
\begin{eqnarray}
\nonumber 
\Lambda^\pm_{-1}&=&R_{\pol}^\pm,\\\nonumber
\Lambda^\pm_0&=&R_{\pol}^\pm R_{\text{in}}^\pm,\\
Q_0&=&R_{\pol}^++R^-_{\text{in}}.
\end{eqnarray}
The sum-product can be eliminated by subtracting terms proportional to $\Lambda^\pm_{N-1},Q_{N-1}$, leaving just:
\begin{eqnarray}
\label{eq:NLp}
\Lambda^+_N&=&\left(R^+_{\act}(N)+\frac{R^+_{KP}R^+_{\act}(N)}{R^+_{KP}(N-1)}\right)\Lambda^+_{N-1}
-\frac{R^+_{KP}(N)R^+_{\act}(N)R^+_{\act}(N-1)}{R^+_{KP}(N-1)}\Lambda^+_{N-2}+R^+_{KP}(N)\frac{R_{\pol}^+}{R_{\pol}^-}\Lambda^-_{N-1},
\\\label{eq:NLm}
\Lambda^-_N&=&\left(R^-_{\act}(N)+R^-_{KP}(N)+\frac{R^-_{KP}(N)R^+_{\act}(N)}{R^-_{KP}(N)}\right)\Lambda^-_{N-1}
-\frac{R^-_{KP}(N)R^+_{\act}(N)R^-_{\act}(N-1)}{R^-_{KP}(N-1)}\Lambda^-_{N-2},
\\
Q_N&=&R^+_{\act}(N)Q_{N-1}+\frac{\Lambda^-_N}{R_{\pol}^-}+(R_{\pol}^+-R^+_{\act}(N))\frac{\Lambda^-_{N-1}}{R_{\pol}^-},
\label{eq:NQ}
\end{eqnarray}
with the same initial conditions as above. This system of recursion relations may be used to generate the terms of the spanning tree sums quickly. 
 
In certain simple cases eqns.~\ref{eq:NLp},~\ref{eq:NLm},~\ref{eq:NQ} can be solved as a function of $N$. For example, when the reaction rates are not a function of $N$, such as:
\begin{eqnarray}
\nonumber
        R^+_{\text{in}}&=&k_1 M_{\text{in}},\\\nonumber
        R^-_{\text{in}}&=&k_{1}e^{-\Delta G_y},\\\nonumber
        R_{\act}^+(n)&=&k_{\act},\\\nonumber
        R_{\act}^-(n)&=&k_{\act}e^{\Delta G_{\act}},\\\nonumber
        R_{KP}^+(n)&=&k_{KP}M_{\act},\\
        R_{KP}^-(n)&=&k_{KP}e^{-\Delta G_y},
        \label{eq:Nrates}
\end{eqnarray}
for $n\in\{1,\cdots N\}$, for the step-wise process with monomer tip $\&xy$, where $k_i$ are some overall rates, $M_{\text{in}}$, $M_{\act}$ represent the concentrations of inactive or active monomers, $\Delta G_{\act}$ represents the chemical work upon moving a monomer up one activation stage. The rates in eqn.~\ref{eq:Nrates} are used for the numeric results in figures~\ref{fig:NLoopGamErr} and \ref{fig:NLoopEfficiency}. In this case, the sums over spanning trees are:
\begin{eqnarray}
\nonumber
\Lambda^+_N(y,x)&=&k_{\pol}e^{-\Delta G_x}\Bigg(k_{\act}^N\Big(k_1M_{\text{in}}-k_{1}M_{\act}+k_{\act}e^{\Delta G_y}M_{\act}(e^{\Delta G_{\act}}-1)\Big)\\\label{eq:Nlamplus}
&+&\frac{k_{KP}M_{\act}e^{-\Delta G_y}}{\Delta}\Bigg[\frac{\left(k_{1}\lambda_++k_{\act}(k_{KP}-k_{1})\right)}{(\lambda_+-k_{\act})^2}\lambda_+^{N+1}-\frac{\left(k_{1}\lambda_-+k_{\act}(k_{KP}-k_{1})\right)}{(\lambda_--k_{\act})^2}\lambda_-^{N+1}\Bigg]\Bigg),\\\label{eq:Nlammin}
\Lambda^-_N(y,x)&=&\frac{k_{\pol}e^{-\Gpol} e^{-\Delta G_y}}{\Delta}\Big[\left(k_{1}\lambda_++k_{\act}(k_{KP}-k_{1})\right)\lambda_+^N-\left(k_{1}\lambda_-+k_{\act}(k_{KP}-k_{1})\right)\lambda_-^N\Big],\\\nonumber
Q_N(y,x)&=&\frac{k_{\pol}e^{-\Delta G_x}}{\Delta}\Big[(k_{1}e^{-\Delta G_y}+k_{\act}-\lambda_-)\lambda_+^N-(k_{1}e^{-\Delta G_y}+k_{\act}-\lambda_+)\lambda_-^N\Big]\\
&+&\frac{e^{-\Delta G_y}}{\Delta}\Big[\left(k_{1}\lambda_++k_{\act}(k_{KP}-k_{1})\right)\lambda_+^N-\left(k_{1}\lambda_-+k_{\act}(k_{KP}-k_{1})\right)\lambda_-^N\Big],
\end{eqnarray}
where 
\begin{eqnarray}
        \lambda_\pm&=&\frac{1}{2}\left(k_{\act}+k_{KP}e^{-\Delta G_y}+k_{\act}e^{\Delta G_{\act}}\pm\Delta\right),\\
        \Delta&=&\sqrt{\left(k_{\act}+k_{KP}e^{-\Delta G_y}+k_{\act}e^{\Delta G_{\act}}\right)^2-4(k_{\act})^2e^{\Delta G_{\act}}}.
\end{eqnarray}
From eqns.~\ref{eq:Nlamplus},~\ref{eq:Nlammin} we may write the stall condition. Noting that, $\Lambda^+(y,x)$ is independent of $\Gpol$ and $\Lambda^-(y,x)$ is proportional to $e^{-\Gpol}$, we may write the stall point as
\begin{eqnarray}
        \Gamma(N)=-\ln\left(\frac{\Lambda_N^+(r,r)}{e^{\Gpol}\Lambda_N^-(r,r)}+\frac{\Lambda_N^+(w,w)}{e^{\Gpol}\Lambda_N^-(w,w)}\right),
\end{eqnarray}
such that the dependence on $\Gpol$ in the logarithm is cancelled out.
\end{document}